\documentclass[twocolumn]{aastex63}
\accepted{to ApJS August 30, 2020}
\shorttitle{YEPS III. SBF of SSPs}
\shortauthors{Chung et al.}

\begin{document}
\title{Yonsei Evolutionary Population Synthesis (YEPS) Model. III. Surface Brightness Fluctuation of Normal and Helium-enhanced Simple Stellar Populations}

\correspondingauthor{Chul Chung, Suk-Jin Yoon}
\email{chulchung@yonsei.ac.kr, sjyoon0691@yonsei.ac.kr}
\author[0000-0001-6812-4542]{Chul Chung}
\affil{Equal First Authors}
\affil{Department of Astronomy, Yonsei University, Seoul 03722, Republic of Korea}
\affil{Center for Galaxy Evolution Research, Yonsei University, Seoul 03722, Republic of Korea}
\author[0000-0002-1842-4325]{Suk-Jin Yoon}
\affil{Equal First Authors}
\affil{Department of Astronomy, Yonsei University, Seoul 03722, Republic of Korea}
\affil{Center for Galaxy Evolution Research, Yonsei University, Seoul 03722, Republic of Korea}
\author[0000-0001-5966-5072]{Hyejeon Cho}
\affil{Department of Astronomy, Yonsei University, Seoul 03722, Republic of Korea}
\author[0000-0002-7957-3877]{Sang-Yoon Lee}
\affil{Center for Galaxy Evolution Research, Yonsei University, Seoul 03722, Republic of Korea}
\author[0000-0002-2210-1238]{Young-Wook Lee}
\affil{Department of Astronomy, Yonsei University, Seoul 03722, Republic of Korea}
\affil{Center for Galaxy Evolution Research, Yonsei University, Seoul 03722, Republic of Korea}

\begin{abstract}
We present an evolutionary population synthesis model of the surface brightness fluctuation (SBF) for normal and He-enriched simple stellar populations (SSPs).
While our SBF model for the normal-He population agrees with other existing models, the He-rich population, containing hotter horizontal-branch stars and brighter red-clump stars than the normal-He population, entails a substantial change in the SBF of SSPs. 
We show that the SBF magnitudes are affected by He-rich populations at least $\sim$0.3~mag even in $I$- and near-IR bands at given colors, from which the SBF-based distances are often derived.
Due to uncertainties both in observations and models, however, the SBFs of Galactic globular clusters and early-type galaxies do not allow verifying the He-enriched model.
We propose that when combined with independent metallicity and age indicators such as ${\rm Mg}_2$ and ${\rm H}\beta$, the UV and optical SBFs can readily detect underlying He-rich populations in unresolved stellar systems at a distance out to $\gtrsim 20$\,Mpc.
A full set of the spectro-photometric and SBF data for SSPs from the Yonsei Evolutionary Population Synthesis (YEPS) model is available for download at http://cosmic.yonsei.ac.kr/YEPS.htm.
\end{abstract}


\section{Introduction}
\label{s1}

The surface brightness fluctuation (SBF) is one of the most important distance indicators for unresolved stellar systems at distances out to $\sim$100 Mpc. 
\citet{1988AJ.....96..807T} first quantified the SBF phenomenon observed in external galaxies and suggested that the strength of fluctuation can be used to constrain the distance to galaxies.
After that, the SBF was adopted as the distance measure that can rival SNe Ia for the distance beyond the regime of RR Lyrae or Cepheid variables ~\citep[e.g.,][]{1991ApJ...373L...1T, 1997ApJ...475..399T,2000ApJ...530..625T, 2001ApJ...559..584A, 2013IAUS..289..304B}. 
Starting with $\sim$\,20 galaxies in the Local Group \citep{1991ApJ...373L...1T}, \citet{1997ApJ...475..399T, 2000ApJ...530..625T, 2001ApJ...546..681T} measured the SBF amplitudes for over 300 galaxies out to $\sim$\,40~Mpc.
The distances of over 130 early-type galaxies in the Virgo and Fornax Clusters have been analyzed by using the SBF technique ~\citep[e.g.,][]{2005ApJS..156..113M, 2005ApJ...634..239C, 2007ApJ...655..144M, 2009ApJ...694..556B}, and the SBF is becoming an increasingly precise distance estimator of elliptical galaxies as well as spiral bulges up to 100 Mpc \citep[e.g.,][]{2001ApJ...550..503J, 2008ApJ...678..168B}.

The SBF is also a useful tool for probing stellar population properties such as age, metallicity, and the existence of hot horizontal-branch (HB) stars for elliptical galaxies ~\citep[e.g.,][]{1995AJ....110..179S, 1996AJ....111..208S, 2005ApJ...634..239C, 2007ApJ...662..940C, 2007ApJ...668..130C, 2011A&A...532A.154C, 2011A&A...534A..35C, 2013A&A...552A.106C, 1998ApJ...505..111J, 2004ApJ...611..270G, 2015ApJ...808...91J, 2005MNRAS.363.1279G}. 
In order to interpret the SBF, theoretical efforts have been made to construct SBF models in the extensive wavelength regimes from ultra-violet \citep[UV;][]{1993ApJ...415L..91W} to optical \citep{1990AJ....100.1416T, 1994ApJ...429..557A, 2001MNRAS.320..193B, 2011A&A...532A.154C, 2011A&A...532A.154C, 1993ApJ...409..530W} and to infrared \citep[IR;][]{2001A&A...376..793M, 2005AJ....130.2625R, 2003AJ....125.2783C, 2005MNRAS.362.1208M, 2006A&A...450..979M, 2009ApJ...700.1247R, 2010MNRAS.403.1213G, 2010ApJ...712..833C}. 
Especially, many studies were done on very hot (hot HB stars) and very bright, cool stars (thermally pulsing asymptotic-giant-branch stars; TP-AGB) that greatly influence the SBF amplitudes at short \citep[e.g., $U$-band;][]{1993ApJ...415L..91W} and long wavelengths \citep[e.g., $K$-band, and IR;][]{2010ApJ...710..421L, 2018ApJ...856..170G}, respectively.

In this regard, the discovery of the He-rich stellar subpopulation within the Milky Way globular clusters (GCs) \citep[e.g.,][]{2004ApJ...612L..25N, 2005ApJ...621L..57L, 2017MNRAS.464.3636M} has a significant impact on the SBF research because the He-rich population contains hotter, brighter stars on their He-burning phase \citep{2011ApJ...740L..45C,  2013ApJ...769L...3C, 2017ApJ...842...91C}. 
The presence of hot HB stars associated with the He-rich population increases the SBF measured in short-wavelength passbands.
Also, red-clump stars, which are metal-rich counterparts of hot He-burning stars \citep[e.g.,][]{2015MNRAS.453.3906L, 2017ApJ...840...98J}, are brighter when He-enhanced and thus effectively change the SBF amplitude by increasing the number of bright stars in a given population. 
These characteristics of the SBF can constrain the presence of He-rich populations in the observed galaxies with known distances.

In this series of papers, with an in-depth consideration on the morphological changes of HBs, we presented Yonsei Evolutionary Population Synthesis (YEPS) model which explored the effect of $\alpha$-elements \citep[][YEPS I]{2013ApJS..204....3C} and He abundance \citep[][YEPS II]{2017ApJ...842...91C} on the integrated LICK/IDS absorption indices and the integrated magnitudes of simple stellar populations (SSPs).
This present paper focuses mainly on the effect of the He-enhanced stellar populations on the UV, optical, and near-IR SBF magnitudes. 
We explore the possibility of detecting the He-enhanced stellar population through a cross-analysis of the SBF magnitudes in multi-passbands from UV to near-IR.

The paper is organized as follows. 
The following section describes how our SBF models are constructed. 
Section 3 presents our model results for the SBFs with different assumptions on stellar parameters. 
Section 4 compares our model predictions with other SBF models and the existing observational data.
In Section 5, we discuss the results and implications of our new models.

\begin{figure*}[htbp]
\centering
\includegraphics[angle=-90,scale=0.8]{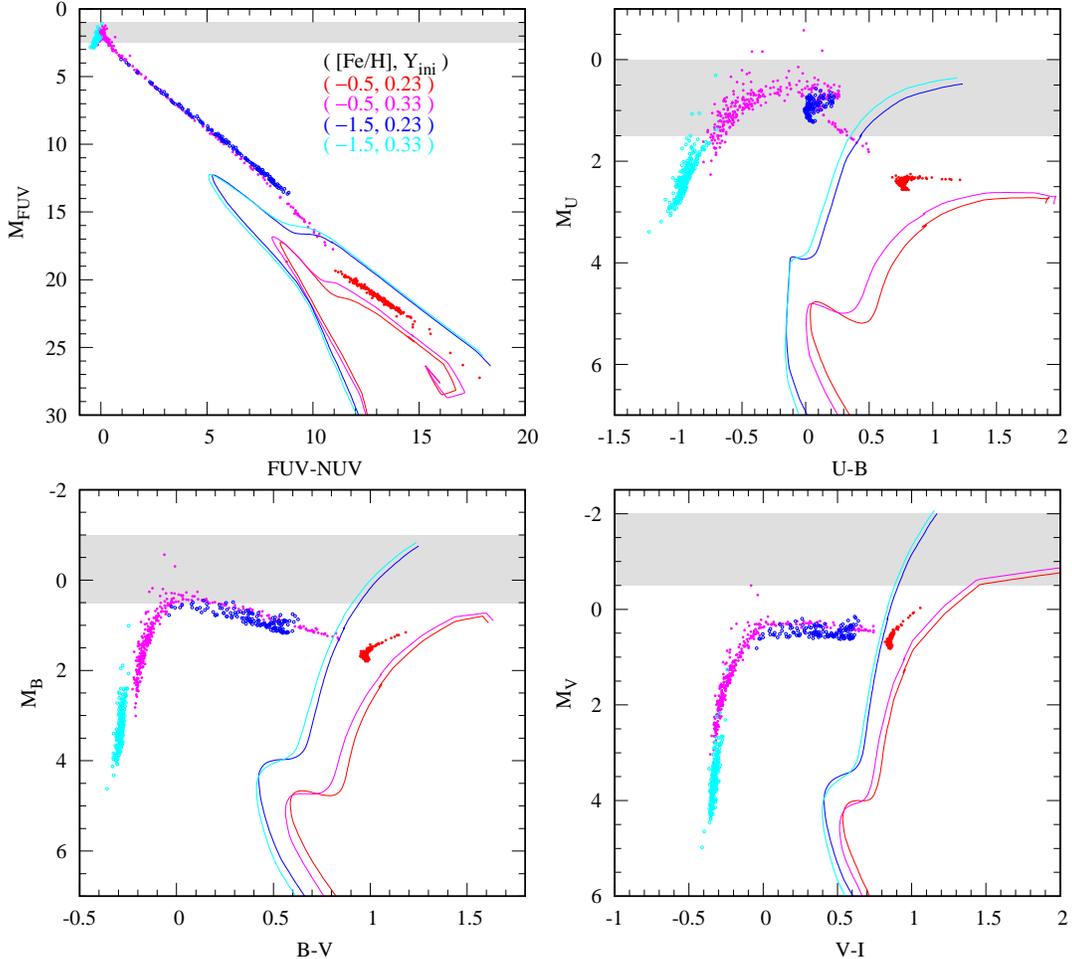}
\caption{CMDs of SSPs with metallicities of ${\rm [Fe/H]}=-1.5$ and $-0.5$ under different assumptions on the initial He of $Y_{\rm ini}=0.23$ and $0.33$. 
The metallicities and initial He are denoted in parentheses in the upper-left panel.
The age of the SSPs is assumed to be 12~Gyr.
The HB stars in the CMDs include the evolution away from the zero-age HB but no observational uncertainty for clarity.
The 1.5~mag wide gray stripe in each CMD indicates a zone where the brightest stars are located at a given bandpass.
\label{f1}}
\end{figure*}

\begin{figure*}[htbp]
\centering
\includegraphics[keepaspectratio,width=\textwidth,height=0.75\textheight]{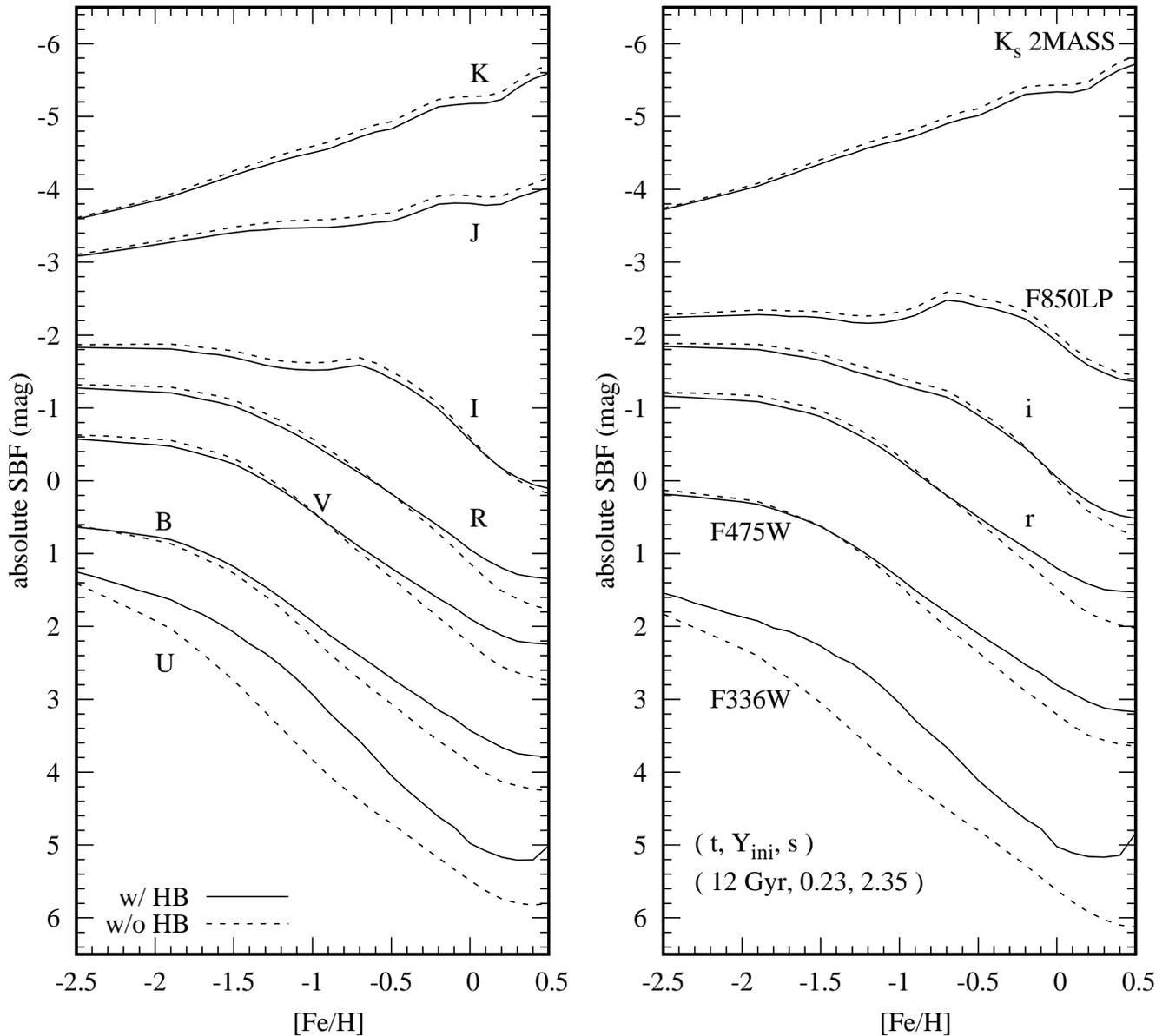}
\caption{YEPS SBF models with and without inclusion of HB stars. 
The solid and dashed lines correspond to the SBF models with and without HB stars, respectively, at the age of 12~Gyr. 
The absolute SBF magnitudes of SSPs in Johnson-Cousins $UBVRIJK$ are presented in the left panel.
The same SBF models for some selected filters on {\it HST}, SDSS, and 2MASS are shown in the right panel.
All SBF magnitudes are given in the Vega mag system.
The model parameters used for the comparison are denoted in the parenthesis in the right panel; $t$, $Y_{\rm ini}$, and $s$ are for age, initial He, and the Salpeter index, respectively.
\label{f2}}
\end{figure*}

\begin{figure*}[htbp]
\centering
\includegraphics[keepaspectratio,width=\textwidth,height=0.75\textheight]{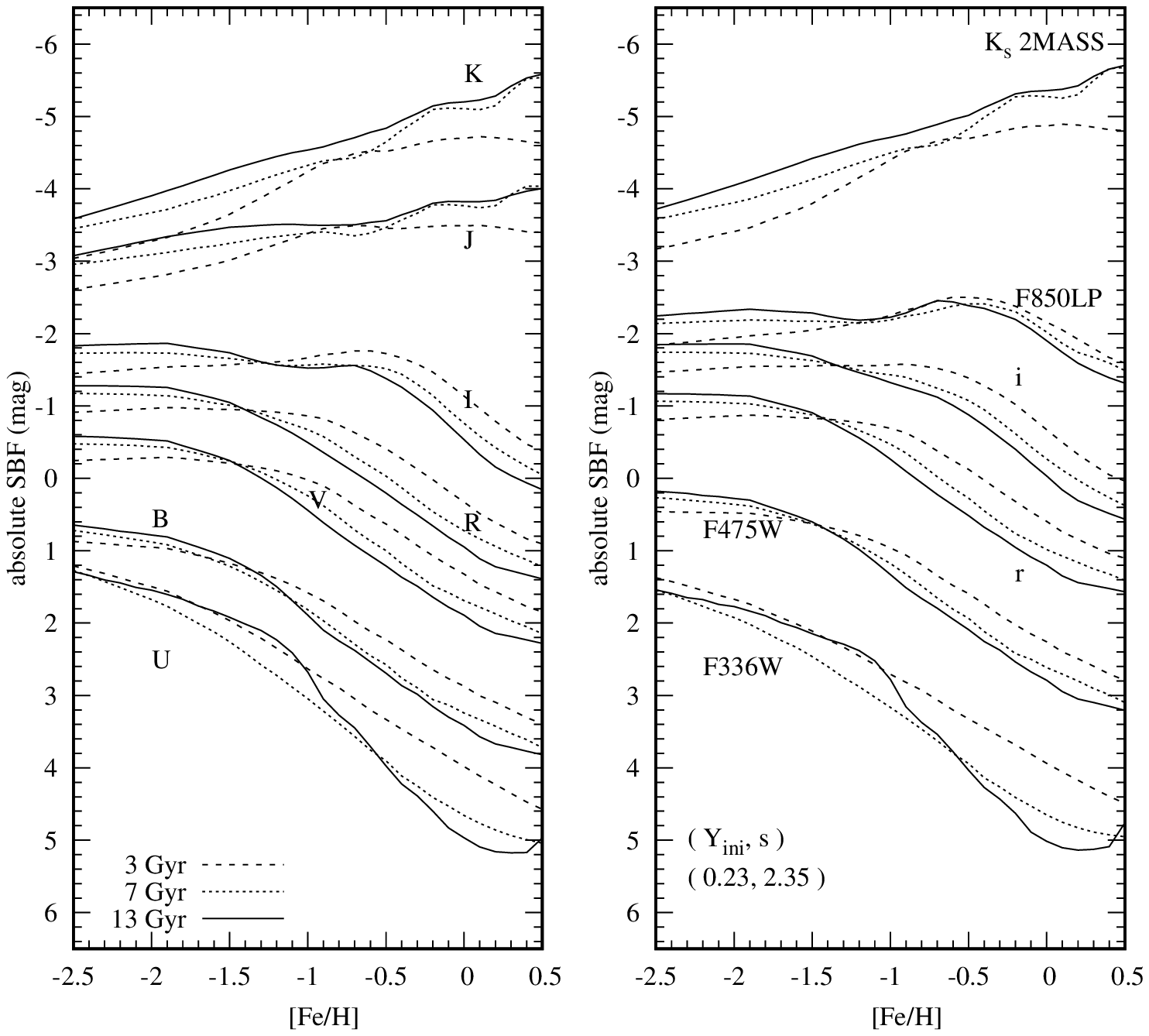}
\caption{Effect of the age on the SBF amplitudes in the same sets of passbands presented in Figure~\ref{f2}.
The dashed, dotted, and solid lines are, respectively, for the SBF models of 3, 7, and 13~Gyr.
The model parameters used for the comparison are denoted in the parenthesis in the right panel.
\label{f3}}
\end{figure*}

\section{Evolutionary Population Synthesis Model Construction}
\label{s2}

The SBF model presented here is {constructed} based on the YEPS for normal-He and He-enhanced populations \citep{2011ApJ...740L..45C, 2013ApJS..204....3C, 2017ApJ...842...91C}. 
All adopted ingredients and input parameters are the same as the model presented in \citet{2017ApJ...842...91C}. 
For a reference, the detailed stellar parameters of the YEPS model are listed in Table 1 of \citet{2017ApJ...842...91C}.
We note that, due to the incomplete carbon-burning stage in the $Y^2$-isochrones \citep{2002ApJS..143..499K}, we employ the BaSTI stellar evolution library \citep{2006ApJ...642..797P}, which has carbon-burning phase mimicking the TP-AGB evolution.
We have constructed additional sets of the SBF model based on the BaSTI library to see the effect of TP-AGB stars on the SBF magnitude.

Figure~\ref{f1} shows synthetic color--magnitude diagrams (CMDs) of 12~Gyr-old SSPs under the different assumptions of metallicity ([Fe/H] = $-$0.5 and $-$1.5) and the initial He abundance ($Y_{\rm ini}$ = 0.23 and 0.33). 
The metal-poor (${\rm [Fe/H]}=-1.5$), normal-He ($Y_{\rm ini}$ = 0.23) population produces HB stars (blue dots) placed between red and blue HB types, while the same metallicity population with richer He ($Y_{\rm ini}$ = 0.33) makes extremely blue HBs (cyan dots).
The red HBs (red dots) are typical of metal-rich (${\rm [Fe/H]}=-0.5$) and normal-He ($Y_{\rm ini}=0.23$) population.  
The He-rich population with $Y_{\rm ini}$ = 0.33, however, produces blue HB stars even for this higher metallicity (magenta dots), which are brighter than other stellar components in $FUV$ and Johnson-Cousins $U$.
The regions where the most luminous populations are located are highlighted with 1.5 mag wide gray shades.
The magnitude of a star at a given passband is functions of temperature and luminosity; as the mean wavelengths of passbands shift to longward, stellar components placed in the gray region changes from hot stars (i.e., blue HBs) to cool stars (i.e., the tip of red giant branches; RGBs).

We utilize the CMDs in Figure~\ref{f1} to calculate the SBF magnitudes in various passbands. 
In order to simulate the SBF ($\overline{L}_{\lambda}$) of SSPs, we adopt \citet{1988AJ.....96..807T} formula which derives the SBF values by dividing the variance of luminosity ($\sum_{i} {n L_i^2}$) by the mean luminosity ($ \sum_{i} {n L_i} = \left< L \right>_{\lambda}$), i.e., 
\begin{equation}
\overline{L}_{\lambda} \equiv \sum_{i} {n L_i^2 / \left< L \right>_{\lambda}} \,,
\label{eq.1}
\end{equation} 
where $n$ is the number of stars in the luminosity $L_{\lambda}$ bins at a chosen passband of $\lambda$.
We calculate SBFs by summing up luminosities of all stars in SSPs and obtain the SBF magnitude at given passbands by applying the bolometric luminosity of the Sun as ${M}_{\odot,\,bol}$ = 4.74.
We use the Vega mag system for SBF magnitudes throughout this paper, except for the comparison shown in Figure~\ref{f12}.

Due to a strong dependency of the SBF on the most luminous stars at given passbands, it is important to avoid the stochastic effect caused by a small number of stars in simulations \citep{2003AJ....125.2783C}. 
Besides, since the influence of hotter and/or brighter HB stars generated from He-rich populations is crucial in this study, it is essential to have enough number of HB stars in simulations. 
Here, we simulated $10^6$ stars at given metallicities and ages of SSPs, which usually produce, on average, $\sim$\,400 and $\sim$\,600 HB stars for $Y_{\rm ini}=0.23$ and 0.33 populations, respectively. 
For post-asymptotic-giant-branch (post-AGB) phase stars, our model employs \citet{1995ApJ...439..875H} stellar evolutionary tracks and has typically two or three stars which are determined  proportionally by the number of HB stars. 
To avoid the rounding up/down of post-AGB stars, we use the time-weighted flux along their evolution tracks. 
While $FUV$- and $NUV$-SBFs are sensitive to the presence of post-AGB stars, these stars exert only marginal or negligible effect on longward of the $U$-band SBF.
The theoretical SBF predictions for SSPs at 12~Gyr are given in Tables~\ref{tab.1}, \ref{tab.2}, and \ref{tab.3} for $Y_{\rm ini}=0.23$, $0.28$, and $0.33$, respectively.

\begin{figure*}[htbp]
\centering
\includegraphics[keepaspectratio,width=\textwidth,height=0.75\textheight]{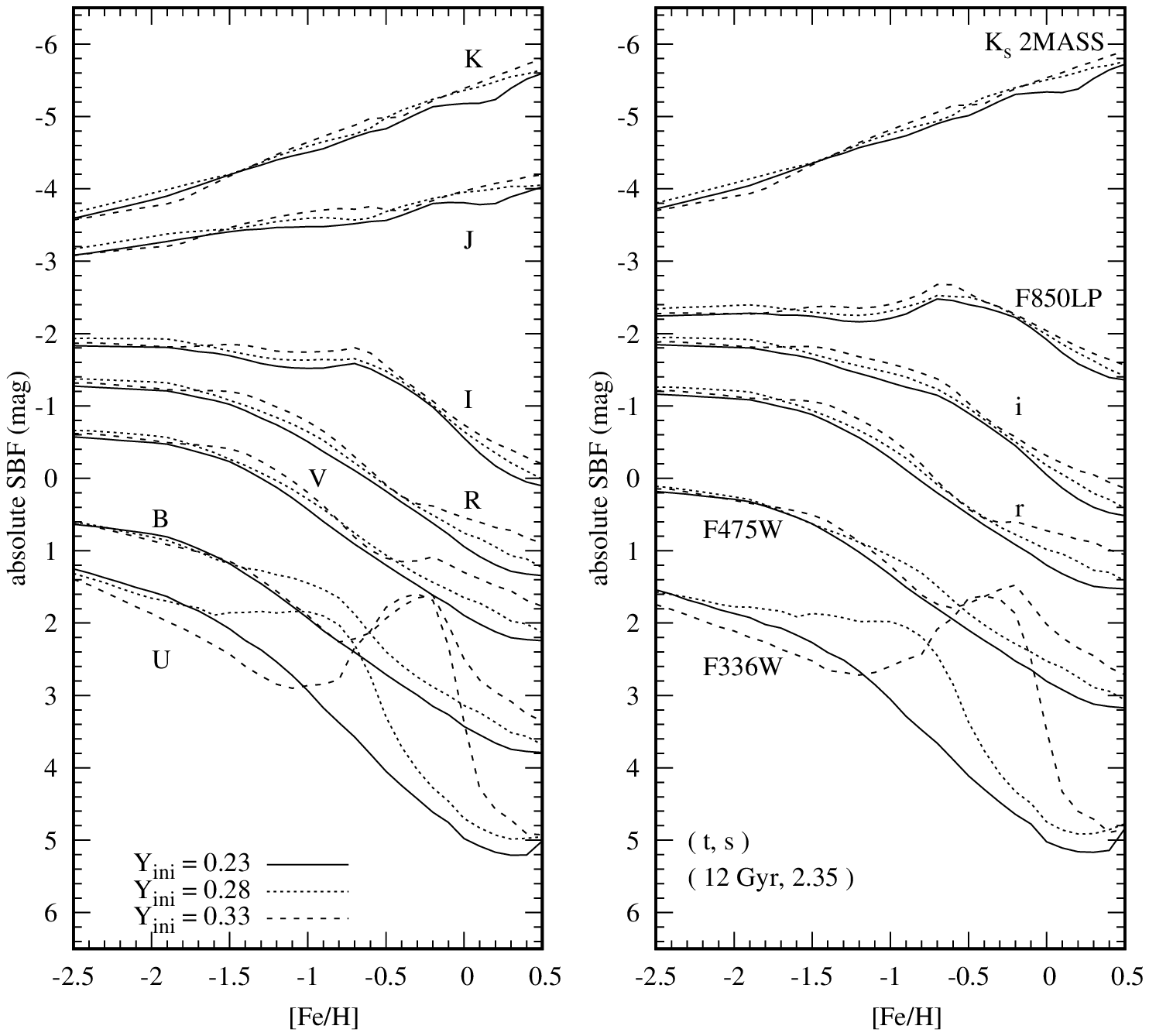}
\caption{Same as Figure~\ref{f3}, but for the effect of initial He on the SBF amplitudes.
The solid, dotted, and dashed lines are, respectively, for the SBF models for $Y_{\rm ini}=0.23$, 0.28, and 0.33.
The model parameters used for the comparison are denoted in the parenthesis in the right panel.
\label{f4}}
\end{figure*}

\begin{figure*}[htbp]
\centering
\includegraphics[keepaspectratio,width=\textwidth,height=0.75\textheight]{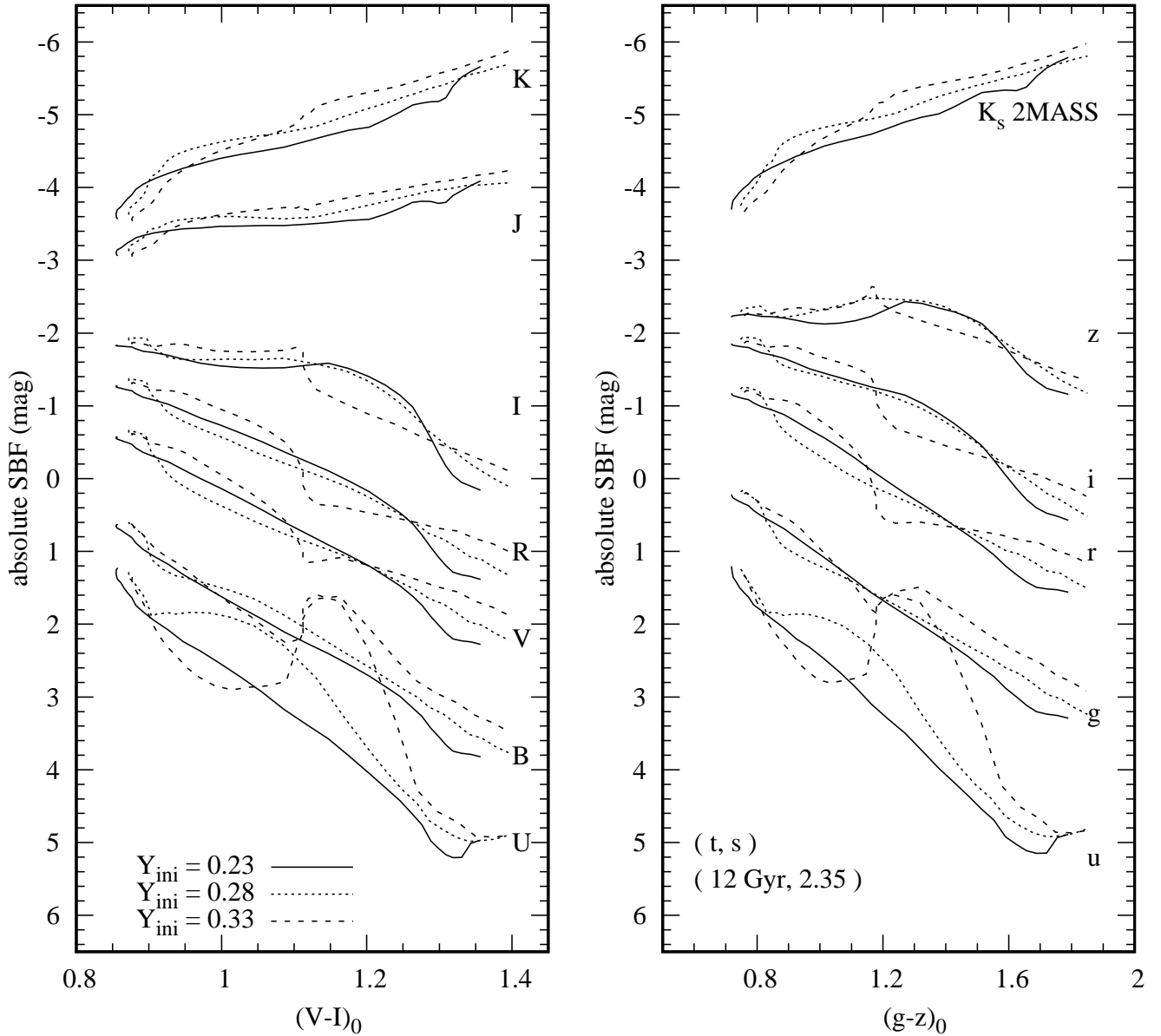}
\caption{YEPS models in the SBF versus integrated color planes with different assumptions on the initial He. 
({\it Left panel}) The same as Figure~\ref{f4} but for the $(V-I)_0$ color. 
({\it Right panel}) SDSS $ugriz$ and 2MASS $K_s$ SBFs as functions of the $(g-z)_0$ color.
All SBF magnitudes are given in the Vega mag system.
The $(V-I)_0$ color is based on the Vega mag system, while the $(g-z)_0$ color is in the AB mag system. 
The model parameters used for the comparison are denoted in the parenthesis in the right panel.
\label{f5}}
\end{figure*}

\begin{figure*}[htbp]
\centering
\includegraphics[keepaspectratio,width=\textwidth,height=0.75\textheight]{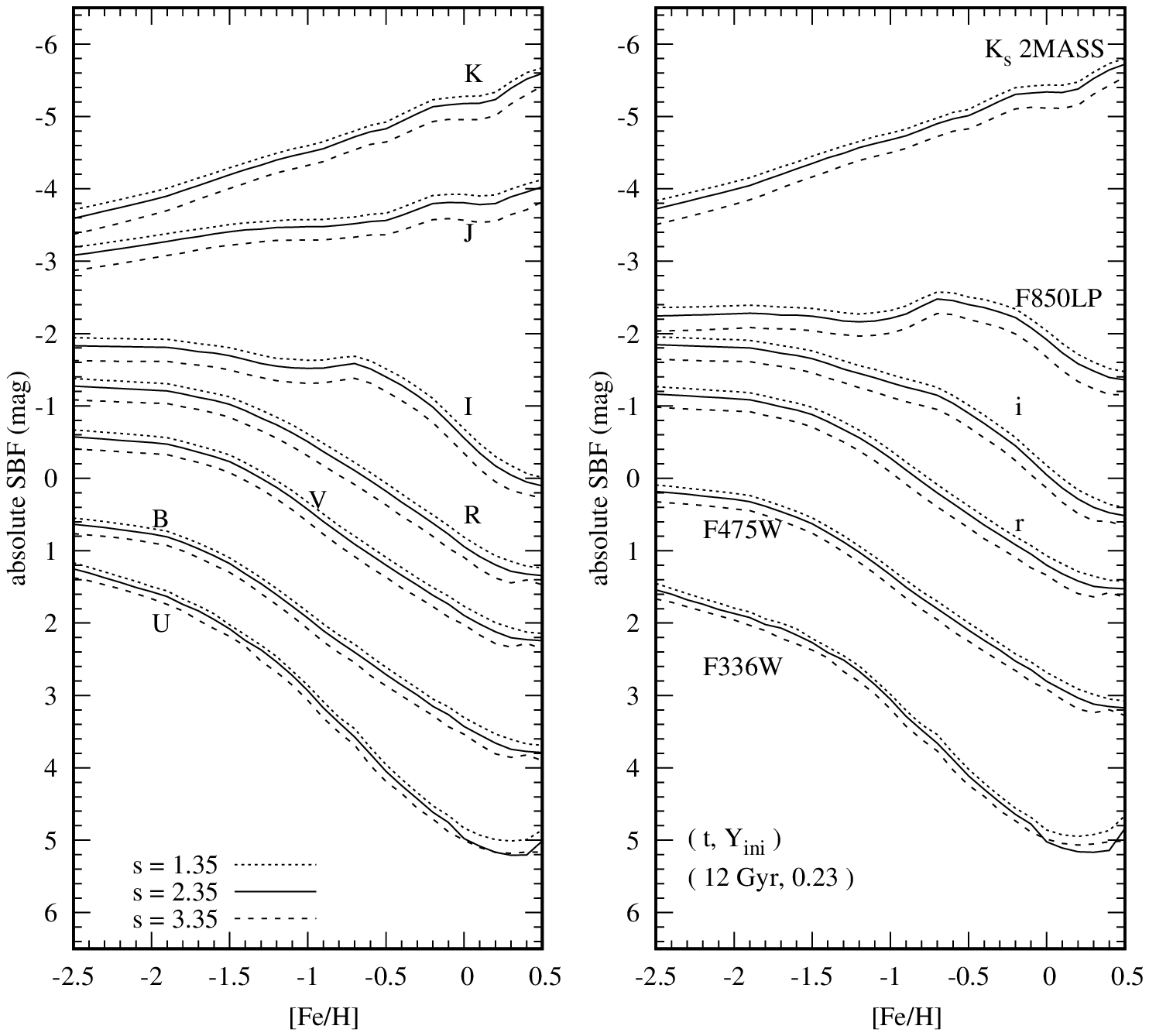}
\caption{Same as Figure~\ref{f3}, but for the effect of the slope of the IMF on the SBF amplitudes.
The dotted, solid, and dashed lines are, respectively, for the adopted Salpeter indices of $s=1.35$, 2.35, and 3.35.
The model parameters used for the comparison are denoted in the parenthesis in the right panel.
\label{f6}}
\end{figure*}

\section{Model Results}
\label{s3}

\subsection{The Effect of HBs on the SBF}
\label{s3.1}

Our SBF models include a thorough treatment of HB stars under the various stellar population parameters, such as age, metallicity, the slope of the initial mass function (IMF), and He abundance. 
Figure~\ref{f2} presents our SBF models with and without the inclusion of HB stars at 12~Gyr. 
The general trend of the SBF amplitudes in $U$- to $I$-bands is that they get fainter as metallicity increases, while near-IR SBFs of $J$- and $K$-bands brighten with increasing metallicity. 
Since the mean luminosity used as the denominator of Equation~\ref{eq.1} overwhelms the variance of luminosity, the $U$- through $I$-band SBFs show decrements at metal-rich regimes.
The decreasing luminosity of the tip of RGBs with increasing metallicity is also part of the reason for this outcome. 
On top of this underlying trend, the effect of HB stars increases the SBF amplitudes in the short-wavelength passbands.
For example, the $U$- and F336W-SBF amplitudes without HBs are fainter than the SBF with HBs in all metallicity regimes. 
As presented back in Figure~\ref{f1}, the mean M$_U$ of HB stars is brighter than all the other stellar components in the metal-rich SSPs (see stellar populations with [Fe/H] = $-$0.5 and $Y_{\rm ini}$ = 0.23). 
The M$_U$ of HB stars in the metal-poor populations is slightly fainter than that of RGB stars, but they are yet one of the most luminous stellar components in the CMD. 
This explains the brighter $U$- and F336W-SBF amplitudes with HBs compared to those without HBs in all metallicity regimes.

As the mean wavelength of passband increases, the intersections of the SBF models with and without HBs moves towards the metal-rich regime. 
From $V$- to $I$-band, the SBFs of metal-poor ([Fe/H] $\leq$ $-$1.0) SSPs with HB are fainter than the SBFs without HB stars. 
The blue HB stars which become fainter in these passbands are responsible for the less bright SBFs (see lower panels of Figure \ref{f1}), and the relative mean luminosity of HB stars at given passbands determines where the intersection points are located.
This trend continues to near-IR bands of $J$ and $K$.

\subsection{The Effect of the Age on the SBF}
\label{theeffectofage}

Figure~\ref{f3} shows the SBFs as functions of [Fe/H] at ages of 3, 7, and 13~Gyr.
The effect of age on the SBF can be interpreted as the close interplay of the three main characteristics of stellar populations, which depend on age. 
The first is the luminosity change of the bright RGBs due to the decreasing mean temperature with age.
The fainter SBFs of older (7 and 13~Gyr) SSPs in the metal-rich regime are the consequence of this effect.
In contrast, the brighter SBFs of older SSPs in the metal-poor regime (${\rm [Fe/H]} \leq -1.5$) are the result of a larger number of RGB stars in the brightest luminosity bin.
The second is a brighter turn-off and red HB stars associated with younger age populations.
This effect explains the brighter SBFs for a 3~Gyr population from $U$ to $I$ in the metal-rich regimes.
The third is hot blue HB stars from the old, metal-poor populations.
This effect causes the almost same or even brighter $U$- and $B$-SBFs of 13~Gyr SSPs in the metal-poor regime compared to 3~Gyr SBF models. 

The near-IR SBFs as a function of age in Figure~\ref{f3} are the expected outcome of the characteristics of the $Y^2$-isochrones employed in this study. 
As presented in \citet{2002ApJS..143..499K} and \citet{2003ApJS..144..259Y}, the tip luminosity of RGBs of the $Y^2$-isochrones is brighter for older ages. 
Besides, unlike the optical magnitude of the tip of RGBs, which are inversely proportional to increasing metallicity, the near-IR magnitude of the tip of RGBs does not show such a trend. 
These effects are combined and explain the age evolution of the SBF amplitudes in the near-IR.
The TP-AGB is also a critical evolutionary stage for young stellar populations with the ages of 1\,$\sim$\,3 Gyr, and has a significant impact on the near-IR SBFs \citep[e.g.,][]{2010ApJ...710..421L}.
A more detailed discussion on this effect is presented in Section~\ref{s3.5}.

\subsection{The Effect of Helium on the SBF}
\label{s3.3}

He-rich stars, in general, evolve faster than normal-He stars.
This leads to a noticeable change in the characteristics of He-burning stars when the age of stellar populations is old enough ($\geq$\,10~Gyr). 
Since the fast evolution reduces the mean stellar mass at given ages, the He-rich population produces hotter HB stars than the normal-He population does.
In addition, the He-rich population, at a given IMF, has more stars in the HB stage.
As shown back in Figure~\ref{f1}, for a He-rich population with $Y_{\rm ini}$ = 0.33, blue HBs are produced even in the metal-rich regime with [Fe/H] $\simeq -0.5$ when the age of the SSP is 12~Gyr, and their number is 50\,\% greater than that of HBs from the normal-He ($Y_{\rm ini}$ = 0.23) population at the same metallicity.

Figure \ref{f4} shows the effect of the He enhancement on the SBF in various passbands. 
The $U$- {and $B$-}SBFs of the $Y_{\rm ini}$ = 0.33 population show great deviation from the SBF with normal-He with $Y_{\rm ini}$ = 0.23 in the metal-rich regime ([Fe/H] $\geq$ $-$1.0).
The $U$- and $B$-SBF magnitudes decrease as the mean temperature of HB stars at a given stellar parameter reaches $T_{\rm eff}$ $\simeq$ 10,000\,K. 
The typical temperature of the brightest blue HBs from $Y_{\rm ini} = 0.33$ population in the $U$ CMD is 10,000\,K (Figure \ref{f1}), and their contribution to the SBF amplitude is maximized when [Fe/H] $\simeq -0.2$. 
The jumps of $U$- and $B$-SBFs for models with $Y_{\rm ini}=0.28$ at ${\rm [Fe/H]} = -1.0$ are also caused by these hot HB stars.
The $U$- and $B$-SBF magnitudes increase rapidly when the temperature of stars is either lower or hotter than 10,000\,K.

The bright SBFs due to blue HB stars seen in $U$- and $B$-bands get weaker at a longer wavelength and almost disappear in the $I$-band. 
The $I$-, $J$-, and $K$-SBFs for the He-rich population, however, show slightly brighter magnitudes than those of the normal-He population in all the metallicity ranges.
This is another manifestation of the fast evolution of He-rich stars, and it makes He-rich stellar populations mimic older ages. 
Although, in these near-IR bands, the luminosity of the brightest RGB stars does not differ between the normal-He and He-rich populations, the number of stars located in the brightest RGBs is larger in the He-rich population compared to the normal-He population. 
These stars are responsible for slightly brighter SBFs in these passbands.

In Figure~\ref{f5}, we present the SBF magnitude versus color relations with respect to the different initial He assumptions.
For the integrated $(g-z)_0$ color, we use the AB mag system for the comparison.
If we use the integrated colors as proxies for metallicity, the effect of enhanced He on SBF magnitudes is more exaggerated at a given color. 
Due to the well-established nonlinear color--metallicity relation of SSPs caused by hot blue HB stars \citep[e.g.,][]{2006Sci...311.1129Y, 2019ApJS..240....2L}, the effects of metallicity and He are combined and result in sizable changes in the SBFs for various passbands. 
For instance, the effect of He-rich populations on the SBFs of relatively longward passbands is still present at $(V-I)_0 \gtrsim 1.1$ and $(g-z)_0 \gtrsim 1.2$.   
At a given $(g-z)_0$ color of $\sim$\,1.3, the SBFs of all SDSS bands and even the 2MASS $K_s$-band change at least $\sim$\,0.4 mag.
We emphasize that, with the distances measured independently, our model could help to detect the presence of the He-rich stellar populations in galaxies.

Figures~\ref{f4} and \ref{f5} show that although the $Y_{\rm ini}=0.28$ population does not produce blue HB stars in the metal-rich regime, it also makes the SBF brighter than in the normal-He populations. 
This may have an implication relevant to the double red-clump stars observed in the Milky Way bulge \citep{2010ApJ...724.1491M, 2010ApJ...721L..28N}.
If the brighter red-clump in the double red-clumps is originated from a He-rich population \citep{2015MNRAS.453.3906L, 2017ApJ...840...98J, 2018ApJ...862L...8L} and if the double red-clump is a common phenomenon in external galaxies like the Milky Way bulge, then this also changes the $U$-, $B$-, $V$-, $R$-, and $I$-SBFs of external galaxies by $\gtrsim 0.3$~mag in the metal-rich regime.

\subsection{The Effect of the IMF slope on the SBF}
\label{theeffectoftheimfslope}

The IMF is a crucial factor that governs the SBF amplitude because it determines the number of bright stars in SSPs. 
Figure \ref{f6} shows the SBF of three different slopes of IMFs under the Salpeter ($s=2.35$), top-heavy ($s=1.35$), and bottom-heavy ($s=3.35$) assumptions at age of 12~Gyr.
Compared to the standard Salpeter IMF, the bottom-heavy IMF increase the SBFs by 0.1\,$\sim$\,0.2 mags and the top-heavy IMF makes the SBFs brighter by $\sim$\,0.1 mag.
The higher fraction of massive, luminous stars for the top-heavy IMF causes the enhanced SBF amplitudes.

It is noteworthy that recent studies suggest that IMFs of early-type galaxies varies with velocity dispersion ($\sigma_V$), $\alpha$-element enhancement, and metallicity \citep[e.g.,][]{2010Natur.468..940V, 2012Natur.484..485C}.
Since the IMF slope change has only a marginal impact on the integrated colors such as $(g-z)_0$ and $(B-I)_0$ \citep[][see their Figure~2]{2013ApJ...769L...3C}, our results present that slight shifts of SBFs originated from the different slope of IMFs are expected at a given metallicity.  

\begin{figure*}[htbp]
\centering
\includegraphics[keepaspectratio,width=\textwidth,height=0.75\textheight]{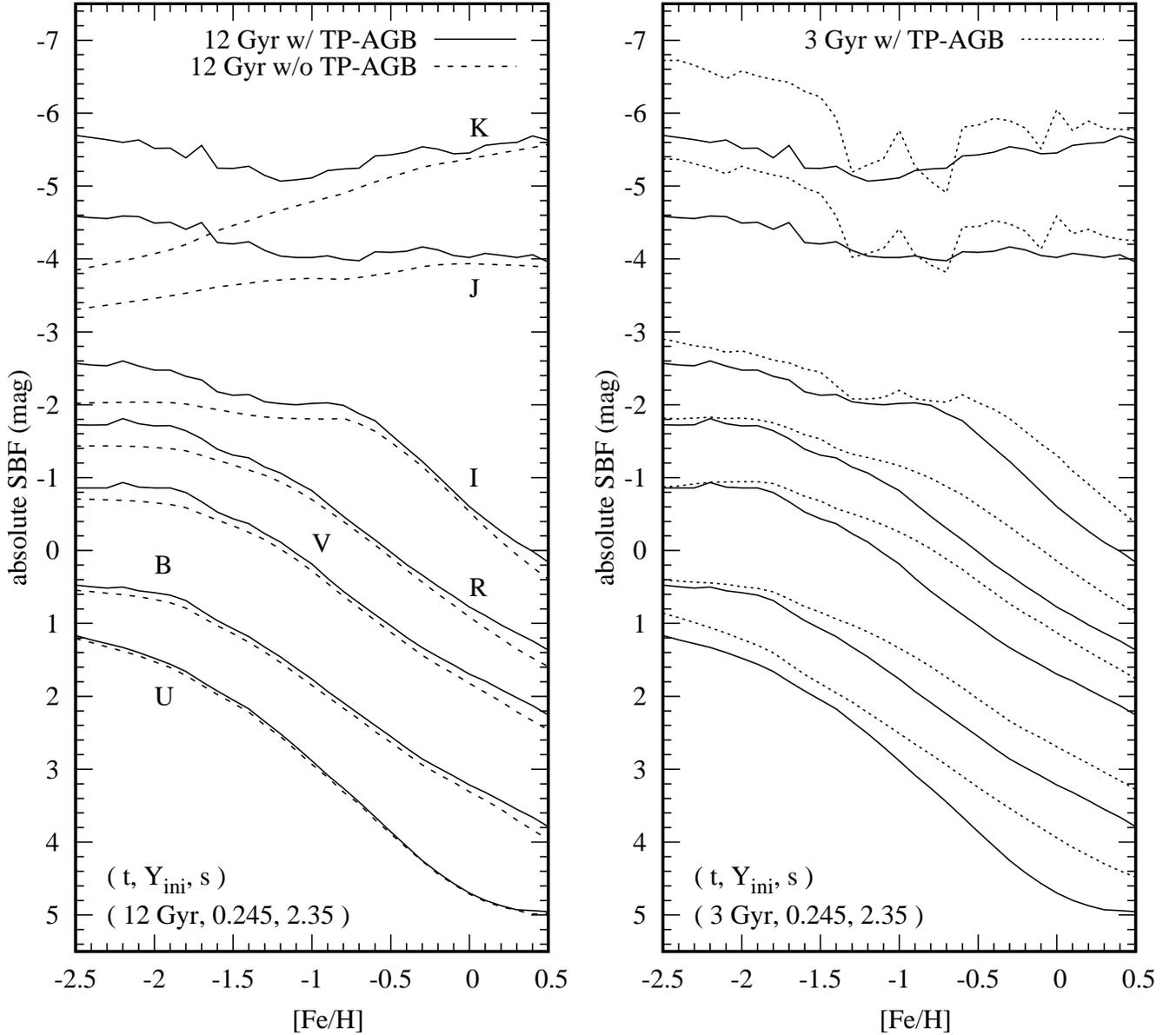}
\caption{SBF models with and without inclusion of TP-AGB stars.
The employed stellar evolution tracks for this model are the BaSTI isochrones \citep{2006ApJ...642..797P}.
All SBF magnitudes are given in the Vega mag system.
The adopted stellar parameters for the model are denoted in the parenthesis at the bottom of each panel.
({\it Left panel}) Solid and dashed lines are the SBF models with and without TP-AGB stars, respectively.
({\it Right panel}) Dotted and solid lines are, respectively, 3 and 12~Gyr models with TP-AGB stars for the same passbands as the left panel.
\label{f7}}
\end{figure*}

\begin{figure*}[htbp]
\centering
\includegraphics[keepaspectratio,width=\textwidth,height=0.75\textheight]{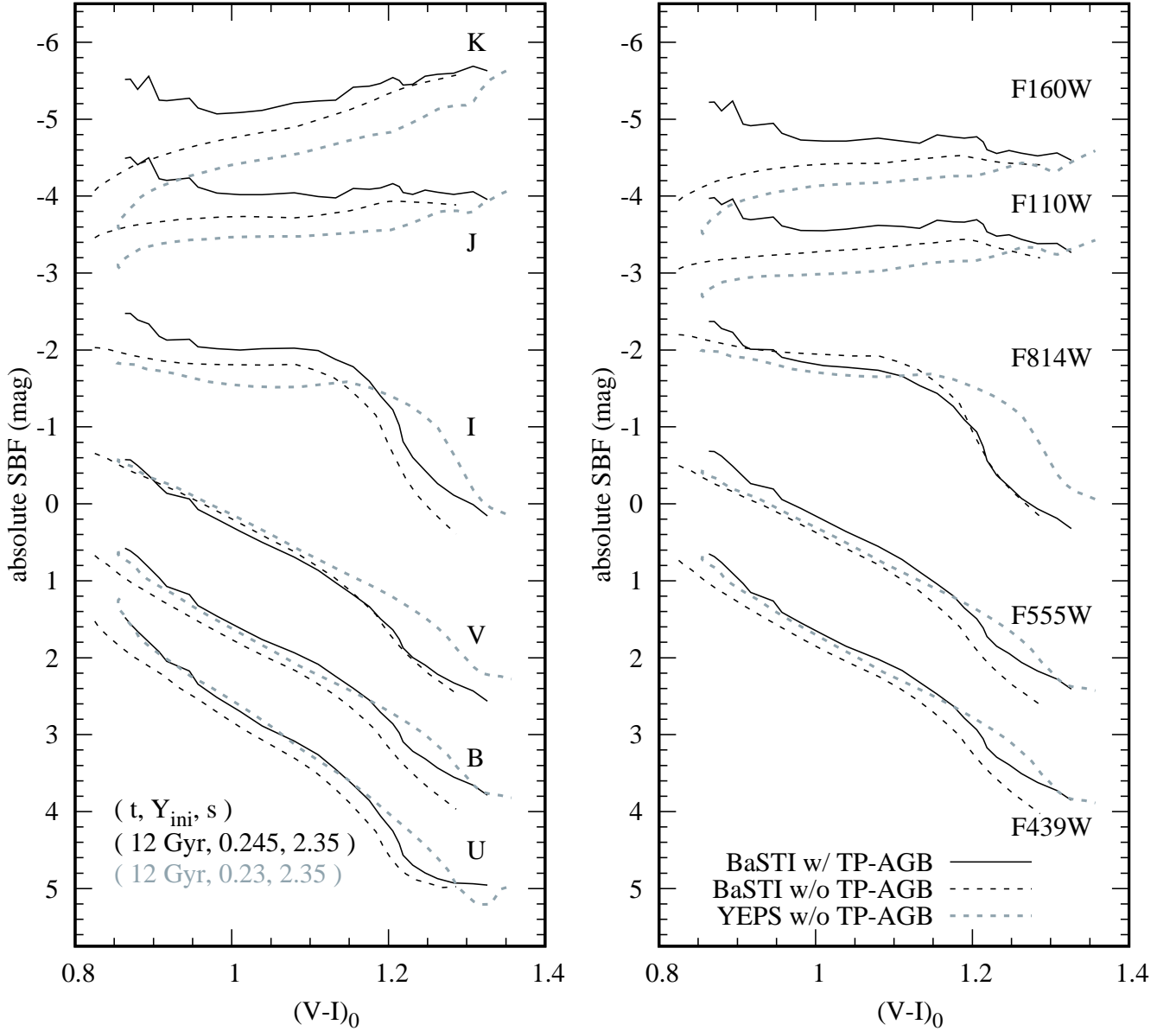}
\caption{Comparison in the SBF versus $(V-I)_0$ planes among the models using the BaSTI isochrones with TP-AGB stars (black solid lines) and without them (black dashed lines), and the canonical YEPS model (gray dashed lines).
All SBF magnitudes are given in the Vega mag system, and the integrated $(V-I)_0$ color is based on the Vega mag system.
({\it Left panel}) The same as the left panel of Figure~\ref{f7}, but with the x-axis of $(V-I)_0$ instead of ${\rm [Fe/H]}$. 
The adopted parameters for each model are denoted in the parenthesis at the bottom.
({\it Right panel}) The same models for the selected {\it HST} filters. 
\label{f8}}
\end{figure*}

\begin{figure*}[htbp]
\centering
\includegraphics[keepaspectratio,width=\textwidth,height=0.75\textheight]{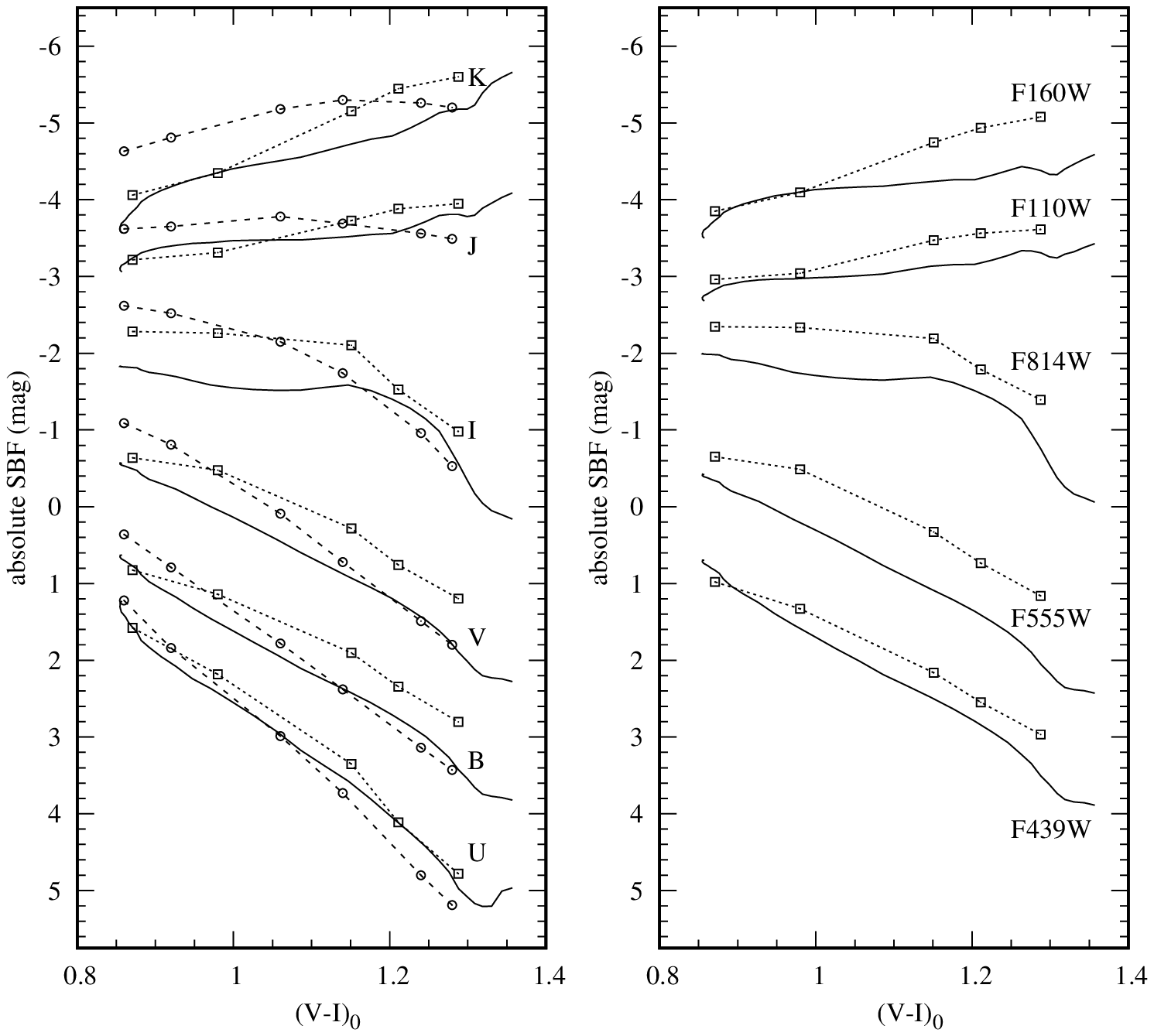}
\caption{Comparison of the YEPS model with \citet{2001MNRAS.320..193B} and \citet{2003AJ....125.2783C} models.
({\it Left panel}) The absolute SBF magnitudes of SSPs in Johnson-Cousins $UBVRIJK$ are presented.
Dashed-circle, dotted-square, and solid lines are, respectively, for the SBF models from \citet{2001MNRAS.320..193B}, \citet{2003AJ....125.2783C}, and YEPS.
To compare similar-age SBFs, the ages of SBF models from \citet{2001MNRAS.320..193B} and \citet{2003AJ....125.2783C} are, respectively, chosen as 12.6 and 13~Gyr, which are equivalent to 12~Gyr of YEPS \citep{2013ApJ...769L...3C}.
The initial He abundance for all models is $Y_{\rm ini}=0.23$.
All models assume the standard Salpeter index for the slope of the IMF.
The SBF magnitudes and the integrated $(V-I)_0$ color are given in the Vega mag system.
Our models are placed in the lower parts of the other two models.
({\it Right panel}) The same as the left panel but for the selected {\it HST} filters. 
\label{f9}}
\end{figure*}

\subsection{The Effect of the TP-AGB on the SBF}
\label{s3.5}

The TP-AGB stars are relatively short-lived but the most luminous stars in the near-IR bands.
Without the appropriate consideration of this evolutionary stage, the SBF model would seriously underestimate the SBF signals in the $I$- and near-IR bands.
The TP-AGB phase is particularly important when the ages of stellar populations are young, ranging from 1 to 3 Gyr, because massive stars ($> 1.5\, M_\odot$) at these ages evolve into the TP-AGB stage \citep[e.g.,][]{1998MNRAS.300..872M, 2003AJ....125.2783C, 2010ApJ...710..421L}.
In order to examine the effect of TP-AGB stars on the SBF, we additionally construct the SBF models based on the BaSTI isochrones.
The simplified treatment of the AGB evolutionary stage in the BaSTI stellar library enables to reproduce the integrated properties of TP-AGB stars in the near-IR bands \citep{2006ApJ...642..797P}.
We use the $\alpha$-element enhanced (${\rm [\alpha/Fe]}=0.4$), normal-He ($Y_{\rm ini}=0.245$) BaSTI library \citep{2006ApJ...642..797P} with metallicities from $Z=0.0001$ to 0.04. 
Other input parameters employed in the SBF model based on the BaSTI isochrones are the same as those used in the YEPS model. 
In our models based on the BaSTI isochrones with $10^6$ stars, the average number of stars in the TP-AGB phase is around 10.

In the left panel of Figure~\ref{f7}, we present the effect of TP-AGB stars on the SBFs in the Johnson-Cousins $UBVRIJK$ photometric system.
Although the quantitative fluctuation caused by the stochastic effect in conjunction with a small number of TP-AGB stars is present, the TP-AGB stars, in general, increases the SBF signals in all tested passbands. 
The impact of TP-AGB stars becomes more robust as the wavelength of measured passbands is shifted longward, and this effect makes even the $K$-band SBF almost $\sim$1.51~mag brighter at the metallicity of ${\rm [Fe/H]} \sim -2.0$.
This trend continues to the $I$-band SBF with $\sim$0.46~mag decrease at the same metallicity.
However, the effect is reduced to $\lesssim 0.13$ when ${\rm [Fe/H]} \sim 0.0$.
Given that the typical metallicity of early-type galaxies is approximately solar metallicity, one should keep in mind this amount of the SBF magnitude shifts when using SBF models without TP-AGB stars.
In Table~\ref{tab.4}, using our model comparison between with and without the inclusion of TP-AGB stars, we summarize the SBF correction terms for normal-He models without TP-AGBs.
To minimize the small number effect of TP-AGB stars, we derive the correction terms based on the seventh-order polynomial fit to the model data.
Since the He-rich stars are less massive than normal-He stars at a given age \citep{2017ApJ...842...91C}, the mean luminosity of He-rich TP-AGB stars is fainter than normal-He counterparts. 
We thus expect smaller correction terms for the He-enhanced SBFs without TP-AGBs.
However, investigating both effects of He-enhancement and the inclusion of TP-AGBs on the SBF magnitude simultaneously is beyond the scope of this paper.
We will fully discuss this issue in our upcoming paper.

In the right panel of Figure~\ref{f7}, we display the effect of the population age on the SBF model with TP-AGB stars.
Unlike the SBFs without TP-AGBs in the left panel of Figure~\ref{f3}, the model shows slightly brighter SBFs in all metallicity ranges.
The effect of TP-AGB stars is strengthened in the metal-poor regimes of near-IR SBFs, causing almost $\gtrsim 3$ mag brighter $K$-band SBF compared to the model without TP-AGB in Figure~\ref{f3}.
We note that, even with the typical metallicity of early-type galaxies ($-0.5 < {\rm [Fe/H]} < 0.5$), the SBF difference induced by the young TP-AGB stars is on average $\sim$0.3~mag in $K$-band.  

Figure~\ref{f8} compares SBF models with and without the inclusion of TP-AGB stars in the integrated $(V-I)_0$ color.
For early-type galaxies at $1.05 \lesssim (V-I)_0 \lesssim 1.25$, the effect of TP-AGB is up to 0.2\,$\sim$\,0.3~mag in $I$-, $J$-, and $K$-bands.
We further examine the effect of the different choice of stellar evolution libraries on the SBF magnitudes in the figure.
For a fair comparison with our models without TP-AGB stars, we simulate the SBF model without TP-AGB stars based on BaSTI isochrones as well.
Although two models employed different stellar evolution libraries, both of SBF models without TP-AGB exert a fairly small effect on the overall shape of the SBF--color relations and agree well with each other.
The small deviations and some offsets in certain metallicity regions are caused by the intrinsic characteristics of the employed stellar libraries.
However, the choice of different stellar evolution libraries causes the systematic shift in color and results in the almost $\sim$1~mag change in the F814W-SBF. 
Therefore, a careful recalibration of models based on the empirical SBF--color relation is needed when analyzing the stellar population of galaxies within those colors of interest (e.g., $(V-I)_0 \sim 1.25$).

\begin{figure*}[htbp]
\centering
\includegraphics[keepaspectratio,width=\textwidth,height=0.75\textheight]{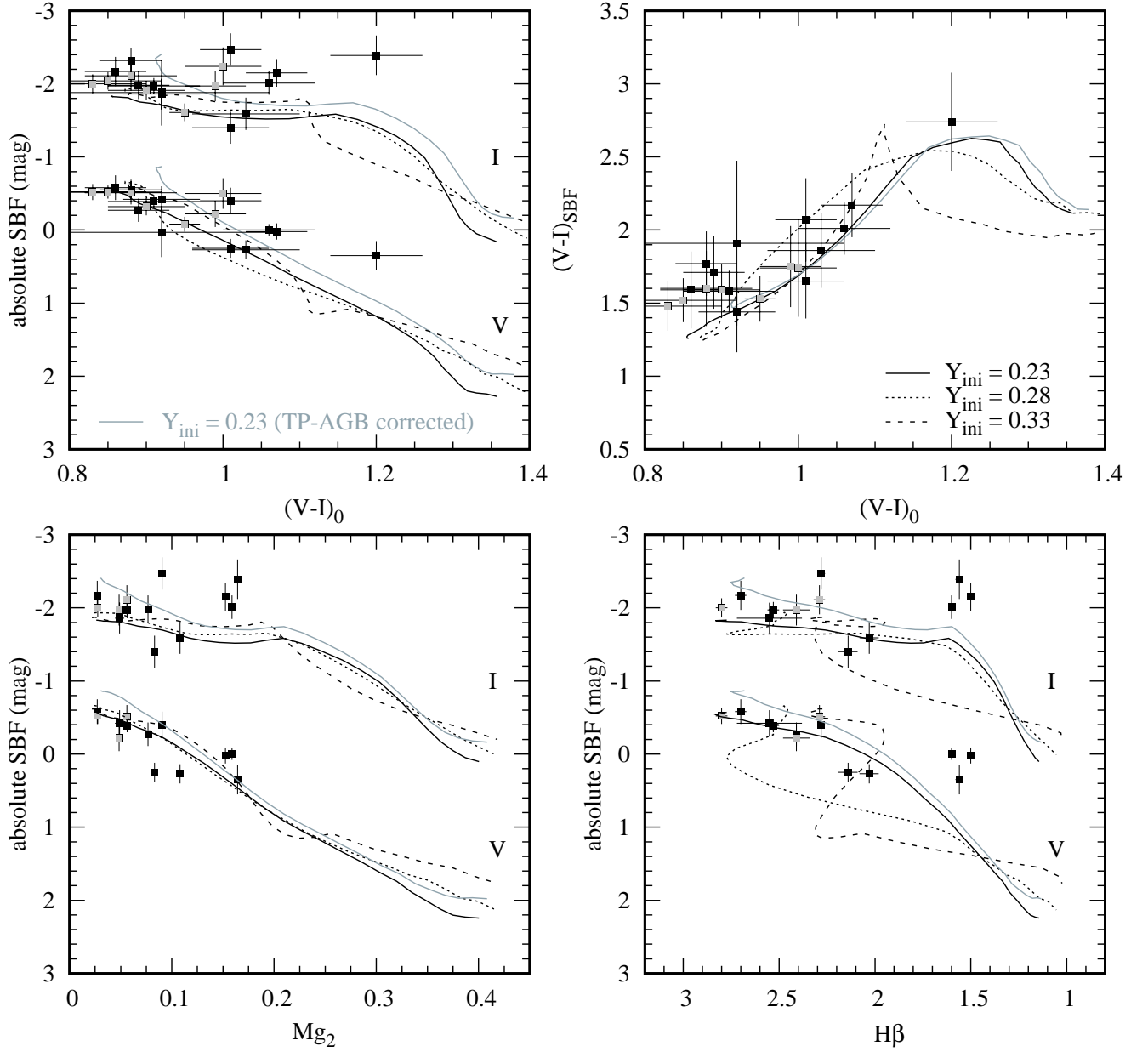}
\caption{Comparisons between the YEPS SBF model and Galactic GC observations.
The upper row shows the absolute $V$- and $I$-{SBF magnitudes} (upper-left panel) and $(V-I)_{\rm SBF}$ (upper-right) as functions of $(V-I)_0$.
The lower row shows the absolute $V$- and $I$-{SBF magnitudes} as a function of ${\rm Mg}_2$ (lower-left) and ${\rm H}\beta$ (lower-right).
The absorption indices of Galactic GCs are obtained from \citet{2016ApJS..227...24K}. 
Different initial He assumptions of $Y_{\rm ini}=0.23$, $0.28$, and $0.33$ are, respectively, shown as solid, dotted, and dashed lines.
The model parameters for age and the slope of the IMF are, respectively, 12~Gyr and $s=2.35$.
All SBF magnitudes and the integrated $(V-I)_0$ color are given in the Vega mag system.
The same $Y_{\rm ini}=0.23$ models after applying the TP-AGB correction are shown in gray lines.  
Squares are the SBFs of Galactic GCs from \citet{1994ApJ...429..557A}.
Gray squares mark Galactic GCs hosting extreme HB stars classified by \citet{2007ApJ...661L..49L}.
\label{f10}}
\end{figure*}

\begin{figure*}[htbp]
\centering
\includegraphics[keepaspectratio,width=\textwidth,height=0.75\textheight]{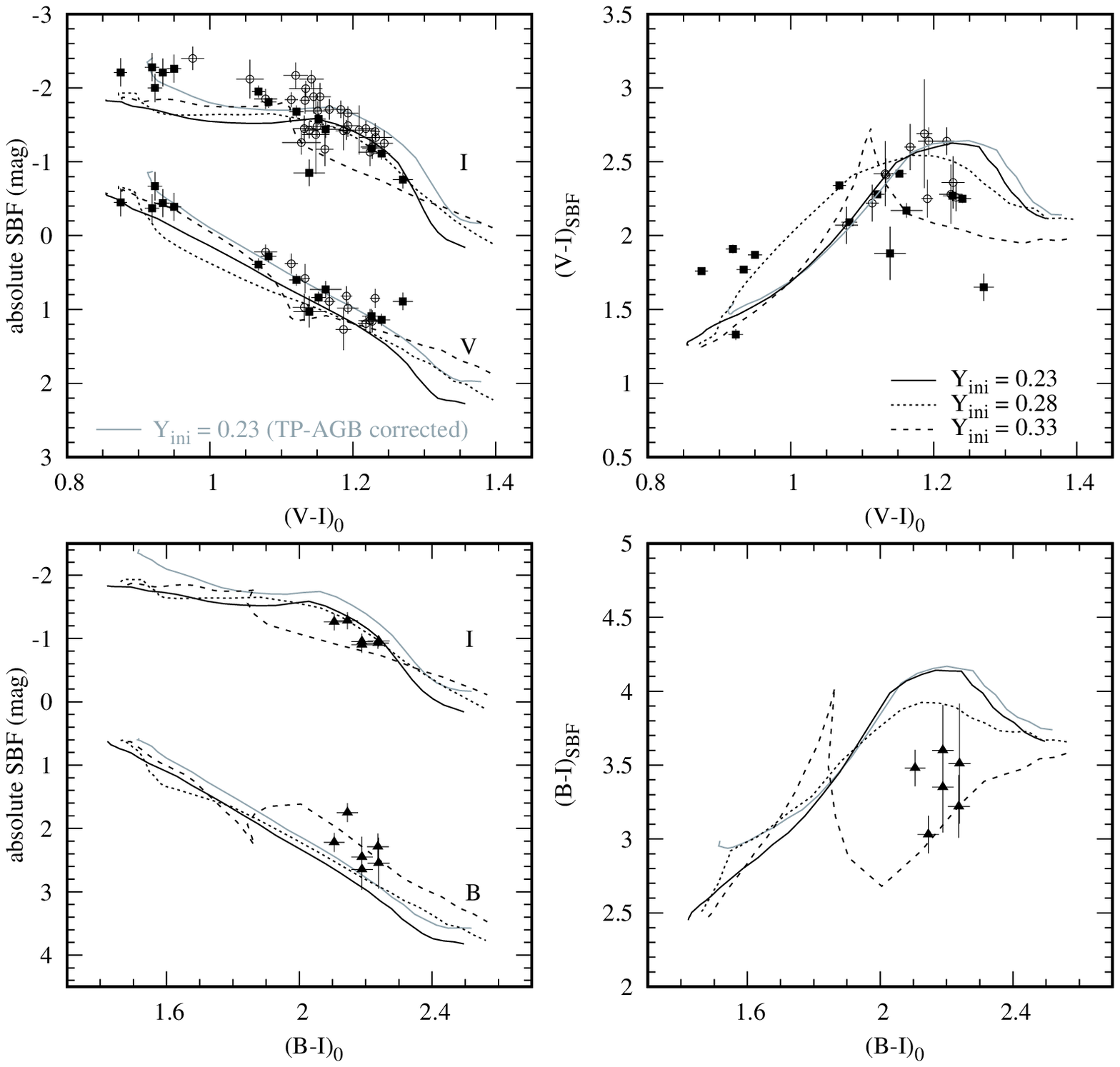}
\caption{({\it Upper panels}) Comparisons between the YEPS SBF model and early-type galaxy observations in $V$ and $I$.
The absolute $V$- and $I$-{SBF magnitudes} (left panel) and $(V-I)_{\rm SBF}$ (right) are shown as functions of $(V-I)_0$.
The SBF models are the same as in Figure~\ref{f10}.
Filled squares are early-type galaxies from the {\it HST} ACS observation by \citet{2007ApJ...668..130C}.
Open circles are early-type galaxies from data compiled by \citet{2003AJ....125.2783C}.
({\it Lower panels}) Same as the upper panels but for $B$ and $I$ SBFs. 
Filled triangles are six early-type galaxies from the {\it HST} ACS observation by \citet{2007ApJ...662..940C}.
\label{f11}}
\end{figure*}

\section{Comparison with other models and observations}
\label{modelcomparison}

\subsection{Comparison with Other Models}
\label{modelcomparisonwothermodels}

Figure~\ref{f9} compares our SBF model with other existing SBF models. 
In the left panel, although all three models were constructed under the different sets of ingredients and assumptions, our model for the normal-He population shows reasonable agreement with the general trend of both \citet{2001MNRAS.320..193B} and \citet{2003AJ....125.2783C} models. 
In particular, the metal-poor part shows a better fit with \citet{2003AJ....125.2783C}, while the metal-rich part follows \citet{2001MNRAS.320..193B} better.
In the right panel, the SBFs of selected {\it HST} passbands are compared with \citet{2003AJ....125.2783C}, showing similar trends with those in Johnson-Cousins passbands.

The SBF amplitude of $I$ at blue $(V-I)_0$ colors ($\lesssim 1.1$) shows values lower than the other models by $\sim 0.5$ mag. 
This is due to the input stellar evolution model of our SBFs that is incomplete in the TP-AGB phase.
As demonstrated back in Figures~\ref{f7} and \ref{f8}, in the metal-poor regime, the TP-AGB stars notably change the $I$- and near-IR SBF magnitudes. 
In addition, the choice of stellar evolution library is also the part of a reason for this offset.
The BaSTI isochrones yield brighter SBFs under the same condition (see Figure~\ref{f8}) compared to the $Y^2$-isochrones. 
These effects seem to cause a deviation of our $I$-SBF from other models. 
The different choice of stellar evolution tracks or the inclusion of TP-AGB stars would improve the fit to the other models for normal-He populations. 

\subsection{Comparison with Observations of the Milky Way Globular Clusters}
\label{modelcomparisonwgalacticgcs}

Figure \ref{f10} presents the comparison of our $V$- and $I$-SBFs at 12~Gyr with the Galactic GCs.
The SBF magnitudes, colors, and distance modulus for each GC are from \citet{1994ApJ...429..557A}. 
In order to demonstrate the differences due to the choice of initial He, we plot models for $Y_{\rm ini}=0.23$, $0.28$, and $0.33$.
In the upper-left panel, the wavy feature appeared in the He-rich model ($Y_{\rm ini}=0.33$) is caused by the different sensitivity to temperature between $(V-I)_0$ and the SBF magnitudes for a given population. 
The $V$-SBF models show reasonable matches to the Galactic GCs within observational uncertainties.
The $I$-SBF models, regardless of the choice of the initial He, predict fainter SBFs compared to the observed GCs.
After we apply the TP-AGB correction (Table~\ref{tab.4}) for $V$- and $I$-SBFs and the integrated color shifts (gray lines), the offset in the $I$-band is much reduced. 
A small number of stars in the observed GCs (${\rm N_{star}} \sim 10^5$) might also be responsible for the offsets of $I$-SBFs.
We mark GCs hosting extreme HB stars originated from He-rich populations \citep{2007ApJ...661L..49L} in gray color and find that the GCs with extreme HBs do not show any particular trend with respect to the normal GCs.
Although the He-rich population decrease SBF magnitudes in the metal-poor regime, the most dramatic effect of the He-rich population on the SBF is expected in $U$- and $B$-bands in the metal-rich region ($(V-I)_0 \gtrsim 1.1$).

In the upper-right panel, we compare our SBF models with the Galactic GCs in the $(V-I)_{\rm SBF}$ color and $(V-I)_{0}$ plane. 
The model in the metal-poor regime appears consistent with the observation within the observational errors.
The fainter $I$-SBF model may be the cause of bluer $(V-I)_{\rm SBF}$ colors compared to some metal-poor GCs with redder $(V-I)_{\rm SBF}$.
Since the TP-AGB corrections for both $V$- and $I$-SBFs make SBFs brighter, the models do not vary much on the SBF color versus the integrated color plane.
The SBFs of $Y_{\rm ini}=0.28$ and $0.33$ show deviation from normal-He SBFs in metal-poor ($0.9 \leq (V-I)_0 \leq 1.1$) and metal-rich ($1.1 \leq (V-I)_0 \leq 1.3$) regime, respectively.
The main driver of these features is hot blue HB stars, which occur at different metallicities according to the initial He. 
If SBFs in shorter bands such as $B$ and $V$ are available for metal-rich GCs with extended blue HB stars (e.g., NGC~6388 and NGC~6441; \citealt{2008ApJ...677.1080Y}), the effect of He-rich populations on the SBF colors can be verified.

In the lower panels, we make the same comparison in the SBF magnitude versus ${\rm Mg}_2$ and ${\rm H}\beta$ diagrams. 
The selected absorption indices are usually used as proxies for metallicity and age of stellar populations. 
We utilize absorption indices of \citet{2016ApJS..227...24K}, which compiled the Lick absorption indices of 53 Galactic GCs. 
Although the matched GCs are mostly located in the metal-poor region, where the He-enhanced SBFs are indistinguishable from the normal-He SBFs, our models show similar trends with the observations.
The SBF measurements of metal-rich Galactic GCs in the short wavelength bands are needed to verify and calibrate our He-rich SBF models.

\begin{figure*}[htbp]
\centering
\includegraphics[keepaspectratio,width=\textwidth,height=0.75\textheight]{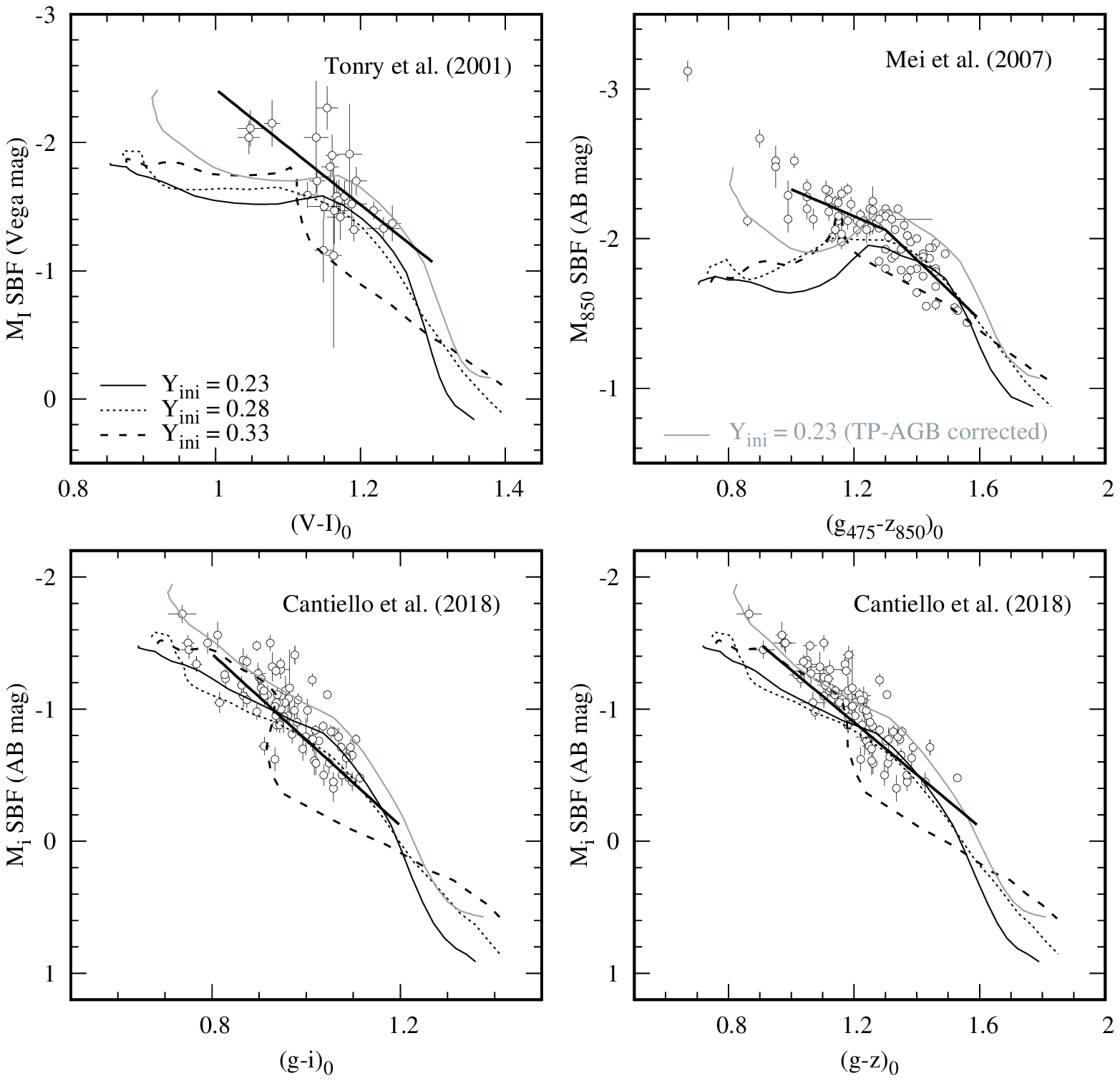}
\caption{Comparing the empirical SBF versus color relations with the YEPS SBF predictions.
Open circles are early-type galaxies in the Virgo Cluster and data are from \citet{2001ApJ...546..681T}, \citet{2007ApJ...655..144M}, and \citet{2018ApJ...856..126C}.
Thick solid lines are empirical SBF--color relations in the literature.
Solid, dotted, and dashed lines are for different assumptions on the initial He of $Y_{\rm ini}=0.23$, 0.28, and 0.33.
The SBF models for $z_{850}$- and $i$-bands are in the AB mag system, and the integrated colors for those bands follow the same photometric system.
The same $Y_{\rm ini}=0.23$ models after applying the TP-AGB correction are shown in gray lines.
The adopted distance modulus for the Virgo Cluster is $(m-M)_{\rm Virgo}=31.09$~mag \citep{2009ApJ...694..556B}.
\label{f12}}
\end{figure*}

\begin{figure*}[htbp]
\centering
\includegraphics[keepaspectratio,width=\textwidth,height=0.75\textheight]{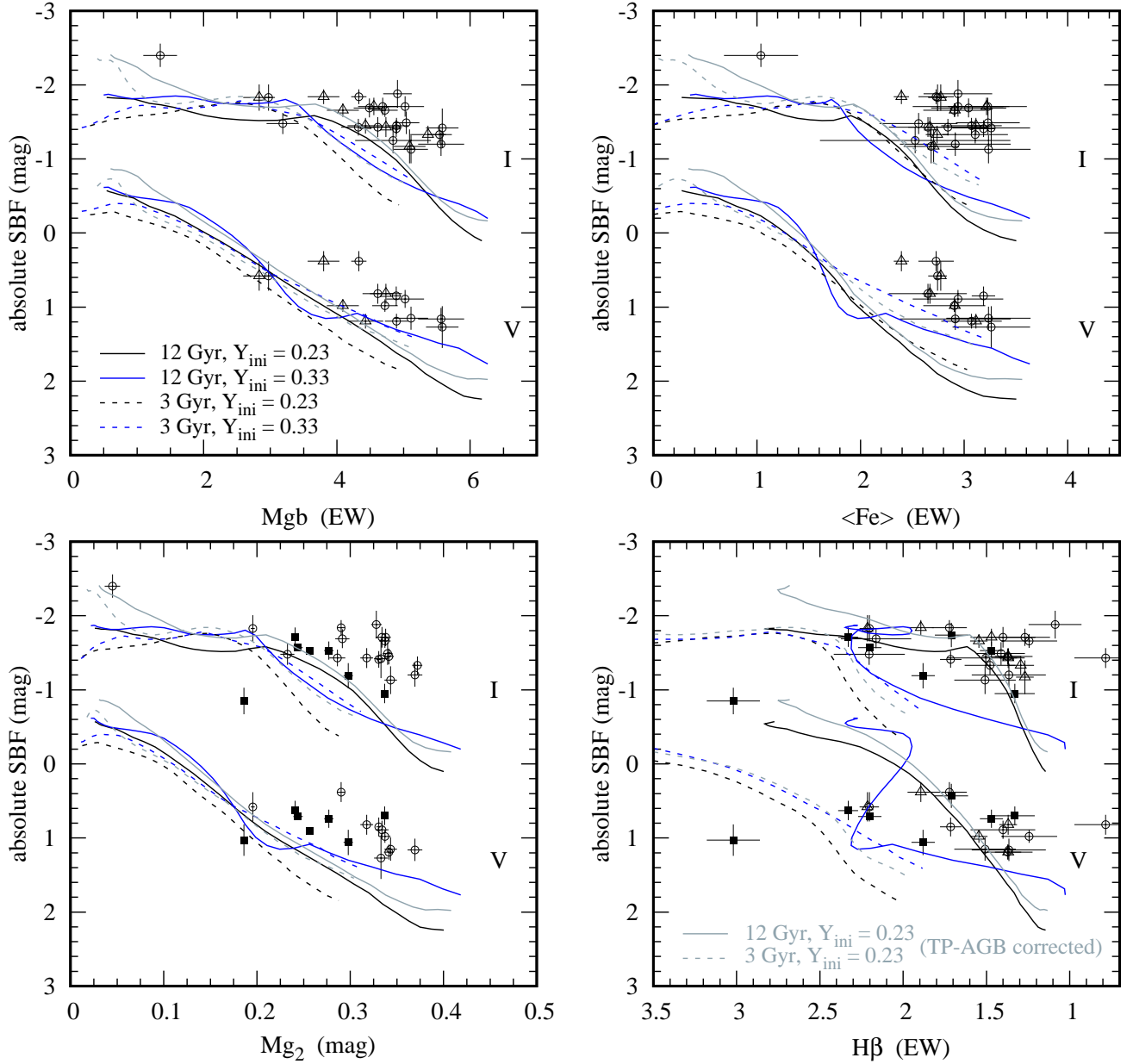}
\caption{Comparisons between the YEPS SBF model and early-type galaxy observations in the absolute SBF mags versus absorption indices planes. 
The absolute $V$- and $I$-SBF mags are shown as functions of ${\rm Mg}b$ (upper-left panel), $\left<{\rm Fe} \right>$ (upper-right), ${\rm Mg}_2$ (lower-left), and ${\rm H}\beta$ (lower-right).
Black and blue solid (dashed) lines are the 12~Gyr (3~Gyr) models for $Y_{\rm ini}=0.23$ and $0.33$, respectively.
Gray solid (dashed) lines are 12~Gyr (3~Gyr) models for $Y_{\rm ini}=0.23$ after applying the TP-AGB correction.
The slope of the IMF is $s=2.35$ for all models.
All SBF magnitudes are in the Vega mag system.
Open symbols are galaxies which have the SBF data from \citet{2003AJ....125.2783C} and the absorption-line data from \citet{2005ApJS..160..176L} {(circles)} and \citet{2006A&A...457..787S} (triangles). 
Filled squares in the lower panels are galaxies with the SBF data from \citet{2007ApJ...662..940C}.
\label{f13}}
\end{figure*}

\subsection{Comparison with Observations of Early-type Galaxies}
\label{modelcomparisonwgalaxies}

The SSP models are rather too simple to represent the whole characteristics of galaxies that consist of composite stellar populations.
Nevertheless, understanding and predicting stellar populations of galaxies based on the `SSP-equivalent' SBF is an attractive part of this study.
In Figure~\ref{f11}, we compare the SBF of early-type galaxies~\citep{2003AJ....125.2783C, 2007ApJ...668..130C, 2007ApJ...662..940C} in $B$, $V$, and $I$ with our SBF models.
Each distance modulus is adopted from the same literature cited above.
In the upper-left panel, the galaxies as a whole follow the TP-AGB corrected model for the normal-He population. 
Given the SBF variation due to the population uncertainties, our model shows fairly good agreements with the observation.
On the other hand, the early-type galaxies in the $B$-SBF (lower-left panel) follow well with the He-rich SBF model. 
As mentioned in Section 3.3, the He-rich population's effect becomes more influential as the mean wavelength of passband decreases.
Hence, this may be an indication of the presence of the He-rich population in these galaxies. 
However, a similar trend is also expected in the younger age model or top-heavy IMF model, so the detailed population analysis is still needed to verify the He-rich population consisting of these galaxies. 
In the right panels, we present the SBF colors as functions of colors.
The effect of the He-rich population significantly changes $(B-I)_{\rm SBF}$ colors.
Some of the galaxies showing stronger $B$-SBFs in the left panel appear to lie closer to the He-rich population model.
Although there exist other parameters affecting the SBF magnitude, given the typical metallicity of early-type galaxies, the SBF in the shorter wavelength passbands can be useful for detecting He-rich populations.

\begin{figure*}[htbp]
\centering
\includegraphics[keepaspectratio,width=\textwidth,height=0.75\textheight]{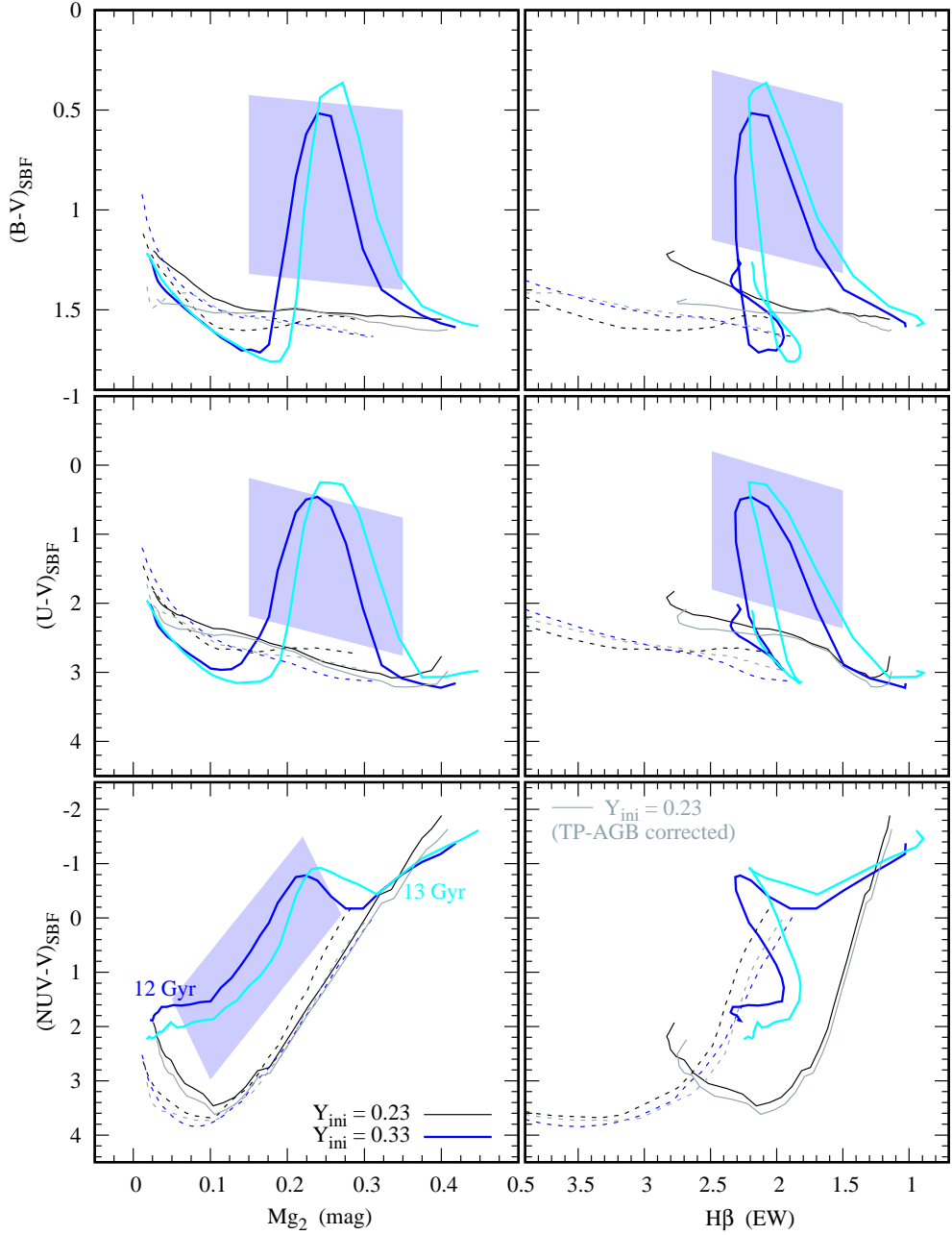}
\caption{Theoretical predictions for SBF colors as functions of absorption-line strengths.
The SBF colors of $(B-V)_{\rm SBF}$, $(U-V)_{\rm SBF}$, and $(NUV-V)_{\rm SBF}$ are presented from top to bottom panels as functions of ${\rm Mg}_2$ (left column) and ${\rm H}\beta$ (right).
Black and blue lines are the models for $Y_{\rm ini} = 0.23$ and $0.33$, respectively. 
Dashed and solid lines are the models for 3 and 12~Gyr, respectively.
Gray lines indicate models for $Y_{\rm ini} = 0.23$ after the TP-AGB correction. 
Cyan lines are the 13~Gyr models for $Y_{\rm ini}=0.33$, showing how older-age and He-rich models behave.
The SBF colors are derived from SBF magnitudes based on the Vega mag system.
Light blue boxes indicate the regions where old, He-rich populations are expected to occupy exclusively. 
\label{f14}}
\end{figure*}

Figure~\ref{f12} presents the comparisons of our models with empirical SBF--color relations defined by the Virgo Cluster galaxies \citep{2001ApJ...546..681T,2007ApJ...655..144M,2018ApJ...856..126C}.
Note that in the upper-left panel, the empirical fit of \citet{2001ApJ...546..681T} is based on galaxies in the Virgo Cluster and other clusters and groups, and open circles represent the Virgo Cluster galaxies only.
The distance modulus for the Virgo Cluster is adopted from \citet{2009ApJ...694..556B} as $(m-M)_{\rm Virgo}=31.09$.
We use the AB mag system for $i$- and $z_{850}$-SBFs.
The integrated colors for $(g_{475}-z_{850})_0$, $(g-i)_0$, and $(g-z)_0$ are also shown in the AB mag system.
Even after the correction for TP-AGBs, there exists some degree of systematic offsets between our normal-He SBF models and the empirical relations for red galaxies with $(g-i)_0 \gtrsim 0.9$ and $(g-z)_0 \gtrsim 1.2$ in the lower panels.
This may be attributed to the incompleteness of our SBF models or the choice of stellar evolutionary library.   
Interestingly, the direction of offsets from TP-AGB corrected models shown in the lower panels is similar to the He enhancement effect of the $0.28 \lesssim Y_{\rm ini} \lesssim 0.33$ populations.
This result is qualitatively consistent with the comparison presented in the lower left panel of Figure~\ref{f11}.
As in the $I$-SBF, the $i$-SBF for $Y_{\rm ini}=0.33$ becomes fainter than normal-He models.
The fainter $i$-SBF of galaxies in the lower panels may show some hint for the presence of the He-rich population.
However, considering that other possible SBF variations originated from different assumptions on the stellar population studied in this paper yield brighter or fainter $i$-SBF\footnote{A higher efficiency of mass loss in the AGB/TP-AGB phase would imply the fainter $i$-band SBF \citep[see e.g.,][]{2003AJ....125.2783C, 2005AJ....130.2625R}.}, it is hard to claim that only the He-rich population can explain this mismatch.
Again, further in-depth studies are needed to confirm the presence of the He-rich populations in early-type galaxies.

In Figure \ref{f13}, we compare our population models with early-type galaxies in the SBF magnitude versus ${\rm Mg}\,b$, $\left< {\rm Fe}\right>$, ${\rm Mg}_2$, and ${\rm H}\beta$ planes.
The early-type galaxy sample is a subsample of \citet{2007ApJ...662..940C}, having observed absorption indices.
We also match early-type galaxies in \citet{2003AJ....125.2783C} with available absorption indices from \citet{2005ApJS..160..176L} and \citet{2006A&A...457..787S}.
We choose ${\rm Mg}\,b$, $\left< {\rm Fe}\right>$, and ${\rm Mg}_2$ as metallicity indicators and ${\rm H}\beta$ as an age indicator. 
Note that SBFs and spectral indices have different radial coverages, and this might cause an extra scatter of data.
The SBF magnitudes versus ${\rm Mg}\,b$, $\left< {\rm Fe}\right>$, and ${\rm Mg}_2$ are similar to the SBF magnitudes versus $(V-I)_0$ in Figures~\ref{f10}-\ref{f12}.
In the ${\rm Mg}\,b$, ${\rm Mg}_2$ and $\left< {\rm Fe} \right>$ versus SBF plane, the He-rich SBF models for different ages do not differ significantly from one another.
By contrast, in the ${\rm H}\beta$ versus SBF, the different age assumptions (3 and 12~Gyr) are well separated from each other. 
One galaxy (NGC~2865) with strong ${\rm H}\beta$ of $\sim$3.0 firmly suggests the existence of younger populations in this galaxy.

For detecting the He-rich population within galaxies, we suggest the most favorable combination of the SBFs and absorption indices, which are least affected by other stellar population parameters.
Figure~\ref{f14} shows the plots of ${\rm Mg}_2$, ${\rm H}\beta$, $(B-V)_{\rm SBF}$, $(U-V)_{\rm SBF}$, and $(NUV-V)_{\rm SBF}$. 
In the top and center panels, the blue shaded areas [$(B-V)_{\rm SBF} \leq 1.3$, $(U-V)_{\rm SBF} \leq 2.5$, $0.15 \leq {\rm Mg}_2 \leq 0.35$, and $1.5 \leq {\rm H}\beta \leq 2.5$] are where the effect of hot HB stars originated from the He-rich population can be maximized, and neither normal-He nor young population cannot enter.
We also show the TP-AGB corrected prediction of our $Y_{\rm ini}=0.23$ SBF models.
The models with TP-AGB stars do not enter the area of interest where the He-rich population can be located.
Considering currently available SBF data in short-wavelengths, the presented $(B-V)_{\rm SBF}$ versus absorption indices are the optimal combination for detecting the presence of the He-rich population in galaxies at a distance out to the Fornax Cluster ($\gtrsim 20$~Mpc; \citealt{2007ApJ...662..940C, 2011A&A...532A.154C, 2013A&A...552A.106C}).
In the lower panels, we demonstrate an area where $(NUV-V)_{\rm SBF}$ identifies the old He-rich population. 
In $(NUV-V)_{\rm SBF}$, the He-rich population would be separated from other populations only by the metallicity indicator ${\rm Mg}_2$, because the ${\rm H}\beta$ models for He-rich populations overlap with the young, normal-He models.
With the more short-wavelengths SBF available, we could have more chances to detect He-rich populations.
The UV-SBF properties of the well-known GCs with He-rich populations can be applied to external galaxies through cross-validation from $NUV$-, $U$-, and $B$-SBFs.
For instance, using the Ultraviolet Imaging Telescope (UVIT) of AstroSat, which is currently in active operation, $NUV_{\rm SBF}$ and optical $V_{\rm SBF}$ observation would be sufficient to identify the possible He-enriched stellar populations in the Milky Way GCs.

\section{discussion}
\label{discussion}

We have presented the YEPS models for the SBF of SSPs for normal-He and He-rich stellar populations.
While our SBF models for normal-He populations agree well with other existing SBF models, SBFs for He-rich populations show substantial changes in $U$-, $B$-, and $V$-SBFs.
Our models predict that the He-rich population has a noticeable impact on SBF magnitudes at given integrated colors up to $\sim$0.5~mag in $V$ and $\sim$3~mag in $U$ compared to the normal-He models at the same condition.
This effect of He-rich populations decreases and almost disappears in the metallicity versus SBF magnitude plane as the SBFs are measured at longer wavelegnths.
However, in color versus SBF magnitude plane, the effect of He-rich populations exists even in the $K$-SBF ($\sim$0.4~mag for $Y_{ini}=0.33$) due to the nonlinear color--metallicity relation.

Another critical implication drawn by SBFs of He-rich populations is that the SBF is a promising He-rich population indicator at UV to the optical regime because the He-rich population is the main formation channel of hotter and brighter HB stars. 
We have proposed that the SBF colors of $(B-V)_{\rm SBF}$, $(U-V)_{\rm SBF}$, and $(NUV-V)_{\rm SBF}$ with absorption indices, such as ${\rm Mg}_2$ and ${\rm H}\beta$, can be used to constrain the presence of the He-rich population in stellar systems. 
Hence, we anticipate that the on-going projects of UVIT \citep[e.g.,][]{2017AJ....154..233S} would provide a good testbed for SBFs of He-rich populations by analyzing Milky Way GCs with multiple stellar populations.
The $NUV$ GC observations of UVIT project can shed light on the SBF detection in the UV regime.

The last but not least noteworthy result is that mildly He-enhanced populations produce cooler and slightly more luminous HB stars in the metal-rich regime \citep[e.g.,][]{2015MNRAS.453.3906L, 2017ApJ...840...98J, 2017ApJ...842...91C} and they also affect SBF magnitudes.     
{We take these stars to be relevant to} the recent discovery of double red-clump stars in the Milky Way bulge \citep[e.g.,][]{2010ApJ...724.1491M, 2010ApJ...721L..28N, 2012A&A...544A.147S}.
The predicted SBF change in $I$ due to mildly He-rich populations is comparable to other SBF model uncertainties originated from the population age, the inclusion of TP-AGB, and the choice of stellar evolution models.    
Given the degree of the SBF variation resulting from the population difference, the proper in-depth analysis of He-rich populations will be possible when combined with the distances measured independently.
In this regard, the relatively short-wavelength SBFs, such as  $B$- and $V$-SBFs, in combination with several absorption indices, are the adequate observables for the required stellar population studies.

\acknowledgments
We thank the anonymous referee for a number of helpful comments. 
C.C., S.-J.Y., and Y.-W.L. acknowledge support provided by the National Research Foundation (NRF) of Korea to the Center for Galaxy Evolution Research (Nos. 2017R1A2B3002919 and 2017R1A5A1070354).
S.-J.Y. acknowledges support by the Mid-career Researcher Program (No. 2019R1A2C3006242) through the NRF of Korea.
H.C. acknowledges support from the Brain Korea 21 Plus Program (21A20131500002).

\bibliographystyle{aasjournal}

\clearpage

\begin{table}
\footnotesize
\begin{center}
\caption{\label{tab.1}Surface brightness fluctuations of YEPS simple stellar population model for $Y_{\rm ini}=0.23$ in the Vega mag.}
\begin{tabular}{ccccccccccccc}
\tableline
Age =   12.0  \\
\tableline
 [Fe/H] & $U$ & $B$ & $V$ & $I$ & $J$ & $K$ & F336W & F475W & F850LP & $i$ & $z$ & $(V-I)_0$ \\        
\tableline
 -2.50 & 1.252 &  0.633 &  -0.571 & -1.831 & -3.081 & -3.592 &  1.540 &  0.180 & -2.244 & -1.844 & -2.230 & 0.854 \\ 
 -2.40 & 1.306 &  0.655 &  -0.556 & -1.826 & -3.109 & -3.640 &  1.601 &  0.200 & -2.248 & -1.836 & -2.233 & 0.854 \\ 
 -2.30 & 1.372 &  0.679 &  -0.542 & -1.823 & -3.141 & -3.690 &  1.677 &  0.220 & -2.254 & -1.830 & -2.237 & 0.856 \\ 
 -2.20 & 1.436 &  0.710 &  -0.525 & -1.820 & -3.173 & -3.741 &  1.739 &  0.245 & -2.260 & -1.823 & -2.242 & 0.861 \\ 
 -2.10 & 1.506 &  0.739 &  -0.509 & -1.817 & -3.206 & -3.793 &  1.812 &  0.267 & -2.266 & -1.816 & -2.247 & 0.865 \\ 
...  & ...  & ...  & ...  & ...  & ...  & ...  & ...  & ...  & ...  & ...  & ...  & ...     \\ 
  0.00 & 4.976 &  3.429 &   1.896 & -0.558 & -3.807 & -5.179 &  5.022 &  2.803 & -1.915 & -0.049 & -1.795 & 1.288 \\ 
  0.10 & 5.082 &  3.543 &   2.012 & -0.347 & -3.781 & -5.179 &  5.108 &  2.919 & -1.733 &  0.133 & -1.603 & 1.299 \\ 
  0.20 & 5.168 &  3.658 &   2.120 & -0.172 & -3.797 & -5.232 &  5.160 &  3.034 & -1.582 &  0.286 & -1.443 & 1.308 \\ 
  0.30 & 5.206 &  3.743 &   2.202 & -0.045 & -3.888 & -5.388 &  5.167 &  3.121 & -1.482 &  0.400 & -1.335 & 1.318 \\ 
  0.40 & 5.203 &  3.774 &   2.229 &  0.051 & -3.957 & -5.517 &  5.142 &  3.153 & -1.393 &  0.475 & -1.240 & 1.330 \\ 
  0.50 & 5.009 &  3.789 &   2.243 &  0.103 & -4.028 & -5.594 &  4.839 &  3.172 & -1.358 &  0.516 & -1.199 & 1.343 \\ 
\tableline
\end{tabular}
\end{center}
\tablecomments{The entire data of Table~\ref{tab.1} are available at http://cosmic.yonsei.ac.kr/YEPS.htm.}
\end{table}

\begin{table}
\footnotesize
\begin{center}
\caption{\label{tab.2}Surface brightness fluctuations of YEPS simple stellar population model for $Y_{\rm ini}=0.28$ in the Vega mag.}
\begin{tabular}{ccccccccccccc}
\tableline
Age =   12.0  \\
\tableline
[Fe/H] & $U$ & $B$ & $V$ & $I$ & $J$ & $K$ & F336W & F475W & F850LP & $i$ & $z$ & $(V-I)_0$ \\        
\tableline
-2.50 & 1.314  & 0.603  & -0.662  & -1.929  & -3.167  & -3.672  &  1.566  & 0.113  & -2.341  & -1.943  & -2.327  &  0.872  \\
-2.40 & 1.376  & 0.633  & -0.658  & -1.935  & -3.202  & -3.722  &  1.605  & 0.129  & -2.354  & -1.946  & -2.339  &  0.873  \\
-2.30 & 1.441  & 0.675  & -0.641  & -1.932  & -3.235  & -3.773  &  1.640  & 0.160  & -2.360  & -1.939  & -2.344  &  0.880  \\
-2.20 & 1.517  & 0.722  & -0.626  & -1.931  & -3.270  & -3.828  &  1.687  & 0.191  & -2.369  & -1.935  & -2.351  &  0.886  \\
-2.10 & 1.593  & 0.768  & -0.608  & -1.930  & -3.305  & -3.881  &  1.738  & 0.223  & -2.377  & -1.929  & -2.359  &  0.891  \\
...  & ...  & ...  & ...  & ...  & ...  & ...  & ...  & ...  & ...  & ...  & ...  & ...     \\
 0.00 & 4.696  & 3.138  &  1.648  & -0.646  & -3.950  & -5.358  &  4.745  & 2.523  & -1.987  & -0.177  & -1.864  &  1.286  \\
 0.10 & 4.826  & 3.236  &  1.722  & -0.495  & -3.971  & -5.408  &  4.859  & 2.614  & -1.853  & -0.056  & -1.722  &  1.305  \\
 0.20 & 4.927  & 3.373  &  1.841  & -0.350  & -4.001  & -5.477  &  4.915  & 2.748  & -1.732  &  0.075  & -1.594  &  1.323  \\
 0.30 & 4.983  & 3.515  &  1.971  & -0.212  & -4.032  & -5.547  &  4.909  & 2.889  & -1.621  &  0.209  & -1.476  &  1.341  \\
 0.40 & 4.975  & 3.567  &  2.013  & -0.101  & -4.032  & -5.585  &  4.855  & 2.941  & -1.504  &  0.291  & -1.355  &  1.358  \\
 0.50 & 4.949  & 3.681  &  2.129  &  0.011  & -4.052  & -5.644  &  4.773  & 3.060  & -1.413  &  0.410  & -1.259  &  1.377  \\
\tableline
\end{tabular}
\end{center}
\tablecomments{The entire data of Table~\ref{tab.2} are available at http://cosmic.yonsei.ac.kr/YEPS.htm.}
\end{table}

\begin{table}
\footnotesize
\begin{center}
\caption{\label{tab.3}Surface brightness fluctuations of YEPS simple stellar population model for $Y_{\rm ini}=0.33$ in the Vega mag.}
\begin{tabular}{ccccccccccccc}
\tableline
Age =   12.0  \\
\tableline
 [Fe/H] & $U$ & $B$ & $V$ & $I$ & $J$ & $K$ & F336W & F475W & F850LP & $i$ & $z$ & $(V-I)_0$ \\        
\tableline
 -2.50 & 1.395 & 0.627 & -0.616 & -1.865 & -3.080 & -3.570 &  1.747 &  0.144 & -2.271 & -1.882 & -2.258 & 0.876 \\ 
 -2.40 & 1.469 & 0.650 & -0.618 & -1.873 & -3.106 & -3.605 &  1.813 &  0.153 & -2.284 & -1.887 & -2.270 & 0.878 \\ 
 -2.30 & 1.569 & 0.702 & -0.596 & -1.863 & -3.127 & -3.645 &  1.887 &  0.189 & -2.283 & -1.875 & -2.268 & 0.883 \\ 
 -2.20 & 1.669 & 0.753 & -0.573 & -1.853 & -3.149 & -3.683 &  1.967 &  0.225 & -2.281 & -1.861 & -2.264 & 0.890 \\ 
 -2.10 & 1.774 & 0.806 & -0.549 & -1.841 & -3.170 & -3.721 &  2.050 &  0.263 & -2.278 & -1.846 & -2.260 & 0.895 \\ 
...  & ...  & ...  & ...  & ...  & ...  & ...  & ...  & ...  & ...  & ...  & ...   & ...    \\ 
  0.00 & 3.369 & 2.496 &  1.299 & -0.738 & -3.970 & -5.386 &  3.480 &  2.018 & -2.038 & -0.310 & -1.919 & 1.241 \\ 
  0.10 & 4.282 & 2.786 &  1.387 & -0.603 & -4.022 & -5.468 &  4.330 &  2.201 & -1.924 & -0.207 & -1.796 & 1.269 \\ 
  0.20 & 4.571 & 2.954 &  1.485 & -0.483 & -4.072 & -5.561 &  4.602 &  2.342 & -1.818 & -0.109 & -1.683 & 1.296 \\ 
  0.30 & 4.722 & 3.082 &  1.555 & -0.392 & -4.111 & -5.636 &  4.707 &  2.452 & -1.722 & -0.047 & -1.582 & 1.324 \\ 
  0.40 & 4.909 & 3.253 &  1.687 & -0.289 & -4.162 & -5.721 &  4.892 &  2.613 & -1.642 &  0.054 & -1.497 & 1.349 \\ 
  0.50 & 4.923 & 3.354 &  1.767 & -0.198 & -4.193 & -5.793 &  4.851 &  2.711 & -1.563 &  0.137 & -1.413 & 1.373 \\ 
\tableline
\end{tabular}
\end{center}
\tablecomments{The entire data of Table~\ref{tab.3} are available at http://cosmic.yonsei.ac.kr/YEPS.htm.}
\end{table}

\begin{table}
\begin{center}
\caption{\label{tab.4}$\Delta$SBF magnitude correction for normal-He models ($Y_{\rm ini}=0.23$) without TP-AGB stars in the Vega mag.}
\begin{tabular}{cccccccc}
\tableline
Age =   12.0  \\
\tableline
 [Fe/H] & $\Delta U$ & $\Delta B$ & $\Delta V$ & $\Delta R$ & $\Delta I$ & $\Delta J$ & $\Delta K$ \\        
\tableline
 -2.00 & -0.059 & -0.107 & -0.219 & -0.314  & -0.462  & -1.090 & -1.509     \\ 
 -1.50 & -0.035 & -0.066 & -0.115 & -0.163  & -0.269  & -0.581 & -0.789     \\ 
 -1.00 & -0.051 & -0.082 & -0.079 & -0.090  & -0.185  & -0.297 & -0.356     \\ 
 -0.50 & -0.029 & -0.085 & -0.092 & -0.086  & -0.124  & -0.271 & -0.294     \\ 
  0.00 &  0.004 & -0.076 & -0.115 & -0.122  & -0.065  & -0.132 & -0.132     \\ 
  0.50 & -0.046 & -0.213 & -0.244 & -0.266  & -0.267  & -0.109 & -0.104     \\ 
\tableline
\end{tabular}
\end{center}
\end{table}

\end{document}